\newcommand{\bleq}{\ifpreprintsty
                   \else
                   \end{multicols}\vspace*{-3.5ex}{\tiny
                   \noindent\begin{tabular}[t]{c|}
                   \parbox{0.493\hsize}{~} \\ \hline \end{tabular}}
                   \fi}
\newcommand{\eleq}{\ifpreprintsty
                   \else
                   {\tiny\hspace*{\fill}\begin{tabular}[t]{|c}\hline
                    \parbox{0.49\hsize}{~} \\
                    \end{tabular}}\vspace*{-2.5ex}\begin{multicols}{2}
                    \fi}
\newcommand{\bcols}{\ifpreprintsty\else\begin{multicols}{2}\fi}
\newcommand{\ecols}{\ifpreprintsty\else\end{multicols}\fi}

 \documentstyle[aps,epsf,multicol,eqsecnum]{revtex} 

\def\aaa#1=#2={\epsffile{#2}}
\def\epsfig#1{\aaa#1=}

\begin{document}
\draft
\title{
Does quasi-long-range order in the two-dimensional $XY$ model really survive
weak random phase fluctuations?
}
\author{Christopher Mudry$^{1,2}$ and Xiao-Gang Wen$^1$}
\address{
$^1$Department of  Physics, Massachusetts Institute of Technology, 77
Massachusetts Avenue, Cambridge, MA 02139, USA.
$^2$Present address, Lyman Laboratory of Physics,
Harvard University, Cambridge, MA 02138, USA.
}
\date{\today}
\maketitle
\begin{abstract}
Effective theories for random critical points are usually
non-unitary, and thus may contain relevant operators with negative 
scaling dimensions. To study the consequences of the existence of
negative dimensional operators, we consider the random-bond $XY$ model.
It has been argued that the $XY$ model on a square lattice, when
weakly perturbed by random phases, 
has a quasi-long-range ordered phase (the random spin wave phase)
at sufficiently low temperatures.
We show that infinitely many relevant perturbations to the
proposed critical action for the random spin wave phase
were omitted in all previous treatments.
The physical origin of these perturbations is intimately related to
the existence of broadly distributed correlation functions.
We find that those relevant perturbations do enter the
Renormalization Group equations, and affect critical behavior.
This raises the possibility that the random $XY$ model has no
quasi-long-range ordered phase and no Kosterlitz-Thouless (KT) 
phase transition.
\end{abstract}
\begin{multicols}{2}

\section{Introduction}
\label{Introduction}

The theory of the Renormalization Group (RG) and, in particular,
the concept of relevant and irrelevant operators 
provides a general and deep understanding of critical points in
clean systems. The physical picture at the heart of the RG theory is
of sufficient generality to have been applied to the study of critical 
points in disordered systems such as random magnetic systems or
the problem of Anderson localization 
\cite{Aharony96,Kardar 1996,Lee85}.

However, effective theories for critical points induced by disorder
(in short, random critical points) are usually non-unitary,
i.e., they contain operators whose scaling dimensions
need not be bounded from below. 
Thus, it is possible for a random critical point
to be endowed with operators whose scaling dimensions are negative. 
This possibility is closely related to the fact that observables
can be very broadly distributed in a critical phenomena
induced by disorder. 

A paradigm of this situation is the problem of
a ``relativistic'' particle moving in two spatial dimensions
in the background of a static but random vector potential, 
in short the random-Dirac-fermion problem
\cite{Ludwig94,Nersesyan94,Bernard95,Chamon196,Mudry96,Chamon96,Kogan96}. 
This system was found to have a line of critical points
such that each critical point contains 
operators with negative scaling dimensions
that carry trivial quantum numbers associated to the symmetries in the problem
\cite{Chamon196,Mudry96}. 
In other words, 
those operators are relevant and may appear in the effective theory.
Now the question is, do the random-Dirac-fermion critical points really
exist or are they destroyed by the relevant operators?
To answer this question, the following issue needs to be addressed.
Do these unusual relevant operators affect the critical points in the 
conventional way,
can we use the standard RG arguments to study the stability of random 
critical points?
One concern is that negative dimensional operators
might have some special properties that would prevent the use
of standard RG arguments. 
For example, whereas the identity operator has zero scaling dimension,
it does not destabilize a critical point 
since it cannot affect the scaling of any correlation functions.  
The question we want to address in this
paper is what happens to a critical point and the corresponding correlation
functions when relevant operators with negative scaling dimensions appear 
in the effective theory.

A second unusual property of the random-Dirac-fermion critical point 
relative to a generic critical point describing a clean system is that
an infinite number of relevant operators appear simultaneously in the
effective theory. The RG flow in the vicinity of the random-Dirac-fermion 
critical point must then involve infinitely many coupled 
equations in which case the issue of its stability becomes much more intricate.

Since there is no general principle ruling out the existence
of an infinite number of relevant operators with negative scaling dimensions
at a random critical point, it is imperative to
reexamine the stability of random critical points
with this possibility in mind. 
Usually, in the literature on random critical points,
only the effects of a finite number of perturbing operators 
are investigated. The properties of the random-Dirac-fermion critical points 
suggest, however, that it is very important to 
study the scaling properties of ``complex'' operators. 
At a random-Dirac-fermion critical point, 
the scaling dimensions of those complex operators are of the form
$h=h_0 n- g n^2$, where $n$ is an integer that characterizes the complexity of
the operator and $g$ characterizes the strength of the randomness.
The trademark of these complex operators is that, whereas they become more
irrelevant the more complex they are (the larger $n$ is) 
in the absence of disorder, this behavior is reversed with disorder.
In fact, since $n$ is unbounded from above, it is seen that
the random-Dirac-fermion critical points ($g>0$) always have an infinite 
number of negative dimensional operators. For weak randomness, only complex
operators with very large $n$ can have negative dimensions. In early
studies of random-Dirac-fermion critical  points, 
the effects of such complex operators
(strongly irrelevant in the absence of disorder but relevant in the
presence of disorder) were not accounted for. 

In this paper we want to gain more insights into
random critical points characterized by
an infinite number of operators with negative scaling dimensions,
elucidate the physical origin of such operators, 
and study how these operators manifest themselves
in the standard RG treatment of the stability of a random critical point.
To this end, we will reexamine the problem of the 
{\it random bond} two-dimensional $XY$ model
\begin{equation}
H_{XY}:=\sum_{\langle ij\rangle} J_{ij}
\left[1-\cos\left(\phi_i-\phi_j-A_{ij}\right)\right].
\label{H}
\end{equation}
Here, the angles $0\leq\phi_i<2\pi$ are defined on the sites $i$ of a
square lattice.
The positive exchange couplings $J_{ij}$ or spin stiffness
and the real phases $A_{ij}$ are
defined on all directed nearest-neighbor pairs of sites $\langle ij\rangle$
and are independently distributed on the links
$\langle ij\rangle$ with variances (mean values) $g_J$ ($J$)
and $g_A$ ($0$), respectively.

Rubinstein et al. have studied the random bond $XY$ model at the Gaussian level
with randomness in the phase only ($g_J=0$, $g_A>0$) \cite{Rubinstein 1983}.
The Gaussian approximation consists in replacing Eq. (\ref{H})
by the continuum limit
\begin{mathletters}
\label{eq: H_G}
\begin{equation}
H_G:=
{J\over2}
\int\! d^2{\bf x}
\sum_{\mu=1}^2
\left[\partial_\mu(\varphi+\Theta)-A_\mu\right]^2,
\label{cal H}
\end{equation}
where $\varphi$ is vortex free whereas $\Theta$ carries vortices.
The probability distribution for the static random vector potential $A_\mu$
is also taken to be Gaussian, 
\begin{equation}
P[A_\mu]:=
{
\exp\left(-{1\over 2g_A}\int d^2{\bf x}\ A^2_\mu\right)
\over
\int{\cal D}[A_\mu]
\exp\left(-{1\over 2g_A}\int d^2{\bf y}\ A^2_\mu\right)
},
\label{prob distr for A}
\end{equation}
\end{mathletters}
where we adopt the summation convention over repeated indices from now on.
Rubinstein et al. argue that the random bond $XY$ model belongs to the same 
universality class as the Gaussian model and they infer from the Gaussian 
model that, for any given but sufficiently small disorder strength $g_A$, 
there exists a line of critical points 
ending at a KT-like transition. In other words, for fixed $g_A$
the KT phase diagram of the pure system \cite{KT transition}
is preserved albeit with scaling exponents depending on $g_A$
\cite{Rubinstein 1983}.
At the heart of their argument is an estimate for the
disorder average of a two-point correlation function for
an operator associated with vortices. The relevance/irrelevance of
this vortex operator controls the KT-like transition.

The manifold of random critical points found by Rubinstein et al.
is quite special. First, these are non-trivial random critical points
since scaling exponents depend both on temperature and disorder strength.
This property should be contrasted with that of a critical point
for which the effect of disorder is fully accounted for by  
irrelevant random perturbations as happens in the two-dimensional
Ising model with weak bond randomness
\cite{Dotsenko-Shankar-Ludwig}. 
Second, each random critical point is exactly soluble, 
all local operators can be listed and their
scaling dimensions can be calculated. 
Third, there is an infinity of operators associated to vortices that
carry negative scaling dimensions.
We want to use the Gaussian approximation to the random bond $XY$ model
as a testing ground to gain some insights about special properties of random
critical points associated to a spectrum of negative scaling
dimensions without lower bound.

We close this introduction by pointing out that,
besides the relevance of the two-dimensional
random $XY$ model to 
magnetic systems with random Dzyaloshinkii-Moriya interactions 
\cite{Rubinstein 1983},
crystal systems on disordered substrates \cite{Cha 1994},
arrays of Josephson junctions with positional disorder
\cite{Granato 1986}, and vortex glasses \cite{Fisher 1991},
the random $XY$ model is also closely related to spectral properties of
two-dimensional Dirac Hamiltonians with random vector potential and
random mass. In turn, random Dirac fermions in two dimensions can be
connected \cite{Mudry 1998} to statistical problems such as the 
random flux-line model in the mixed phase of superconductors
\cite{Hatano 1996} and driven random diffusion model \cite{Fisher 1984}.
We hope that a better understanding of the random $XY$ model might be useful
to this class of problems.

\section{Results}
\label{Results}

It is easy to show that,
if vortices are not allowed (the spin wave approximation),
the continuum model Eq. (\ref{eq: H_G})
is at a fixed point where the spin $\exp({\rm i}\varphi)$
has algebraic quasi-long-range correlations
for any temperature and any disorder strength.
This phase, the {\it random spin wave phase},
is an exactly soluble random critical point.

In this paper we would
like to reexamine the stability of this quasi-long-range ordered phase.
The stability of this phase has been studied before.
If one assumes that the ground state in the vortex sector
is in the dipole phase and
if one considers the binding and un-binding of the simplest
vortices within the first non-trivial order of
a fugacity expansion for the vortices
\cite{Rubinstein 1983}, one finds that the
random critical point is stable only for a range of temperatures and disorder
strengths bounded by the dashed line in Fig. 1.
A more sophisticated approximation consists in treating a non-interacting gas
of dipoles in the presence of disorder non-perturbatively in the vortex
fugacity, in which case quasi-long-range order is present in the shaded area 
in Fig. 1 \cite{Fertig-Nattermann-Scheidl-Tang}.
However, in this paper we find that:
\begin{enumerate}

\item 
For any temperature and any disorder strength of the random phases,
randomness in the vortex fugacity generates
an infinite number of relevant terms (most of them carry negative scaling
dimensions) in the critical action of the
{\it Gaussian approximation to the random bond $XY$ model}
that describes the random spin wave phase.

\item
If the continuum model Eq. (\ref{eq: H_G})
is deduced from a random phase only ($g_J=0$, $g_A>0$)
$XY$ model on a lattice, then its vortex fugacity is necessarily random.

\item
Most importantly, the above relevant terms in the critical action 
describing the random spin wave phase do enter the perturbative
RG equations to each order in the fugacity expansion, 
and completely modify the RG flow at long distances. 
Thus, the relevant terms have the potential 
to cause an instability of the random spin wave phase.

\end{enumerate}
Let us call the simultaneous presence of
relevant terms in a critical action and in the RG equations
{\it a perturbative instability}.
Thus, we may say that the random spin wave phase of the
random bond $XY$ model has a perturbative instability
for any temperature and disorder strength,
if the vortex fugacity is itself random.

We would like to stress that when there are finitely many relevant terms,
the perturbative instability implies the instability of the critical
point. Since they enter the RG equations,
the finite number of relevant perturbations either destroys the
algebraic long-range correlations altogether or changes critical exponents.

However, in our case, the perturbative instability corresponds to
infinitely many relevant terms appearing simultaneously
in the critical action and in the RG equations.
It is thus not completely clear to us what are the effects of
such a perturbative instability, after one sums up an infinite number of 
contributions to the RG equations from all relevant operators. 
By contrast, if all but a finite number of relevant operators can be switched
off from the critical action, these remaining perturbations would completely 
alter the correlation functions at long distances.

At the very least, the perturbative instability in the random bond $XY$ model
represents a new situation with regard to the issue of the stability of
random critical points which must be addressed. The possibility of this new 
situation (i.e., the presence of an infinite number of relevant operators
in the critical action)
is closely related to the fact that the moments of a random variable,
say the exponentiated energy of a dipole of vortices,
need not be bounded if the random variable is sufficiently broadly distributed.
Hence, random critical points need not be described by unitary field theories 
and scaling exponents need not be bounded from below. 

Korshunov \cite{Korshunov 1993} was the first to argue that
there might not be any quasi-long-range ordered phase in the
random phase $XY$ model. To this end,
he introduced a sequence of local operators, $O_{r;N}({\bf x})$,
for the replicated random phase $XY$ model labeled by the two index
$r$ and $N$. For given integer values of $r$ and $N\leq r$,
$O_{r;N}({\bf x})$ creates $N$ vortices,
each belonging to a different replica of the $XY$ model,
on site ${\bf x}$. Here, $r$ is the total number of replica.
Korshunov found that for any strength of disorder, the scaling dimension of
$O_{r;N}({\bf x})$ becomes negative upon analytical continuation to the
$r\downarrow0$ limit, provided $N$ is fixed and chosen sufficiently large.

We recall that the replica approach identifies 
a given physical operator $Q$ with a family or sequence of
operators $Q_r$ labeled by the total number of replicas $r$.
In this paper, we identify the family of operators $O_{r;N}$  
(labeled by $r$) studied by Korshunov with a physical operator. 
More precisely, we show how the family of operators $O_{r;N}$ can be
induced in a physical way in the effective action describing the random
spin wave phase.

In the absence of vortices, the disorder average over the two-point function
of a local operator has the form $A|{\bf x}-{\bf y}|^\alpha$ 
in the random bond $XY$ model. On the one hand,
if the fugacity expansion is valid (as is the case in the clean $XY$ model), 
then the exponent $\alpha$ and the coefficient $A$ depend
on the fugacity $Y$ and have an analytic expansion around $Y=0$.
On the other hand, the breakdown of the fugacity expansion can have
three consequences:
\begin{enumerate}
\item
\label{enum1}
$A(Y)$ is not analytic around $Y=0$, but $\alpha(Y)$ is.
\item
\label{enum2}
Both $A(Y)$ and $\alpha(Y)$ are not analytic around $Y=0$.
\item
\label{enum3}
The critical behavior is completely changed or destroyed by the inclusion of
vortices.
\end{enumerate}

By extending the Renormalization Group (RG) 
equations to fourth order in the fugacity,
we can show that the fugacity expansion breaks down
according to scenario 2, i.e., {\it both} the coefficient $A(Y)$ and 
the scaling exponent $\alpha(Y)$ are not analytic functions of the vortex 
fugacity. Furthermore, if conventional RG arguments apply, one may then
conclude that vortices change or destroy the critical line of 
the the random $XY$ model for any temperature and any impurity strength
according to scenario 3.

The paper is organized as follows.
We first show in section \ref{Broadly distributed correlation functions}
that the two-point correlation function studied in
\cite{Rubinstein 1983} to construct the phase diagram has a very
broad probability distribution, a fact expressed in the
underlying critical theory describing the random spin wave phase
by the presence of infinitely many negative dimensional operators.

We then show that the random spin wave phase is perturbatively unstable.
Indeed, this instability manifests itself by the non-analyticity of
the fugacity expansion for disorder averaged correlation functions
due to the existence of infinitely many negative dimensional operators
in the underlying critical theory. A summary of our arguments is presented
in section \ref{Non-analyticity of the fugacity expansion}
with technicalities relegated to appendix \ref{Fugacity expansion}.

Another signature of the perturbative instability, as we show
in section \ref{The perturbative instability},
is illustrated by
the fact that any randomness in the spin stiffness of the random bond
$XY$ model induces infinitely many relevant perturbations to the
critical theory describing the random spin wave phase in the Gaussian
approximation.

The relationship between the random bond $XY$ model, the random bond
Villain model, and the Gaussian approximation is discussed in section 
\ref{The Villain approximation}.
It is pointed out that the fugacity expansion already breaks down
in the random bond Villain model.

Our conclusions are followed up by two appendices.
The first one discusses ground state properties and some probability
distributions for correlation functions are calculated. 
The second one presents a detailed derivation of the perturbative RG analysis
up to fourth order in the fugacity expansion. There it is also shown
that there exists a one to one correspondence between correlation functions 
for the perturbations induced by a random fugacity within the replica approach
of section \ref{The perturbative instability}, 
and the contributions to the fugacity expansion of section
\ref{Non-analyticity of the fugacity expansion}.

\section{Broadly distributed correlation functions}
\label{Broadly distributed correlation functions}

In this section we construct the random spin wave phase and show that it is
described by a manifold of random critical points. The central quantity of
interest in the spin wave phase is the thermal correlation function for two
spins. We characterize uniquely its probability distribution by all its moments.
All moments are well-behaved. 

We then turn our attention to the vortex sector and, in particular, to
the exponentiated energy of a pair of vortices of opposite charges (a dipole) 
in the background of the random vector potential. All moments of this 
exponentiated energy are calculated in  Eq. (\ref{parabola}). 
This is the central result of this paper. 
In contrast to the spin wave sector, higher moments dominate the lower ones.
This property is nothing but the signature of a log-normal
distribution for the exponentiated dipole energy.
Correspondingly, the energy of a dipole is Gaussian distributed in model 
(\ref{eq: H_G}). Conversely,  
the critical theory describing the random spin wave
phase {\it must} contain infinitely many operators with negative scaling 
dimensions that are associated with vortices
in order to account for the Gaussian distribution of the dipole energy.

\subsection{Factorization into a spin wave and a vortex sector}
\label{subsec:Factorization into a spin wave and a vortex sector}

We begin with the model in the continuum defined by
Eq. (\ref{cal H}).
We {\it reiterate} that $A_\mu$ is assumed Gaussian distributed
with variance $g_A$, i.e.,
that 
\begin{equation}
P[A_\mu]\propto 
\exp\left(-{1\over 2g_A}\int d^2{\bf x}\ A^2_\mu\right).
\label{P[A_mu]}
\end{equation}
The justification for this assumption is that the precise shape 
of the probability distribution should leave critical properties unchanged
as long as the probability distribution preserves the short-range nature 
of spatial correlations in the disorder.
In particular, the tails of the probability distribution of {\it non-compact
support} in Eq. (\ref{P[A_mu]}) should not affect critical properties. 

To understand the role of the random vector potential, it is
convenient to decompose it into transverse and longitudinal components:
\begin{equation}
A_\mu=\tilde\partial_\mu\theta+\partial_\mu\eta, 
\quad\tilde\partial_\mu:=\epsilon_{\mu\nu}\partial_\nu.
\label{A-theta,eta}
\end{equation}
The advantage of this decomposition is that 
the partition function derived from 
$H_G=\int d^2 {\bf x}{\cal H}_G$ in Eq. (\ref{cal H}) 
and the probability distribution in Eq. (\ref{P[A_mu]})
both factorize since 
\begin{eqnarray}
&&
{\cal H}_G\!=\!
{J\over2}
\{[\partial_\mu(\varphi-\eta)]^2\!+ 
[\tilde\partial_\mu(\tilde\Theta-\theta)]^2\},\!
\label{factorized cal H}
\\
&&
P[\theta,\eta]\propto
\exp\left\{
-{1\over 2g_A}\int d^2{\bf x}
\left[(\partial_\mu\theta)^2+(\partial_\mu\eta)^2\right]
\right\}.
\label{P[theta,eta]}
\end{eqnarray}
Here, $\tilde\Theta$ is dual to $\Theta$ 
\footnote{
Given $\Theta$, the dual $\tilde\Theta$ is defined by
$\partial_\mu\Theta=\tilde\partial_\mu\tilde\Theta$.
For example, if 
$\Theta=$
$\sum_{i=1}^M m_i\arctan\left[{({\bf x-x}_i)_2\over({\bf x-x}_i)_1}\right]$,
then
$\tilde\Theta$
$=-\sum_{i=1}^M m_i\ln\left|{{\bf x-x}_i\over l_0}\right|$,
where the vorticities $m_i$ are integer and $l_0$ is an arbitrary length
scale.
},
and one must implement
the constraint on the disorder that there be no zero modes:
\begin{equation}
\int d^2{\bf x}\ \theta({\bf x})=0,\quad
\int d^2{\bf x}\ \eta(  {\bf x})=0.
\end{equation}
The ambiguity in the choice of the integration constant 
in Eq. (\ref{A-theta,eta}) is thus removed.

\subsection{The spin wave sector}
\label{The spin wave sector}

The consequences of the random vector potential on the spin wave
sector are trivial. One can shift spin wave integration variables
to 
\begin{equation}
\varphi':=\varphi-\eta.
\end{equation} 
In the absence of randomness in the spin stiffness, 
all correlation functions for
$\exp({\rm i}\varphi)$=$\exp({\rm i}\varphi'+{\rm i}\eta)$
can be calculated. In turn, the disorder average over $\eta$ can be performed
since the probability distribution for $\eta$ is that of a free scalar field
in two dimensions. 

For example, 
\begin{eqnarray}
\overline{
\langle e^{{\rm i}\varphi({\bf y}_1)}\ e^{-{\rm i}\varphi({\bf y}_2)}\rangle^q
}\ &&\propto\
{
\overline{
e^{{\rm i}q\eta({\bf y}_1)}\ e^{-{\rm i}q\eta({\bf y}_2)}}
\over|{\bf y}_1-{\bf y}_2|^{{q\over2\pi K}}}
\nonumber\\
\ &&=\
{1\over|{\bf y}_1-{\bf y}_2|^{{q+g_AKq^2\over2\pi K}}}.
\label{spin wave exponents}
\end{eqnarray}
Thermal averaging is denoted by angular brackets. Disorder averaging
is denoted by an overline and induces a quadratic dependency on the 
moment $q$ for the scaling exponent.
Thus, the impact of the quenched random
vector potential on the spin wave sector is to drive the system to a new
critical point for any strength of the disorder $g_A$ and for any
reduced spin stiffness 
\begin{equation}
K:=J/T.
\end{equation}
The random vector potential is seen to destroy the
long-range order at vanishing temperature by replacing it with quasi-long-range
order. For all finite temperatures, the algebraic decays of the spin
correlation functions are more pronounced due to the disorder.
On the other hand, random spin stiffness remains irrelevant since it amounts
to a random temperature
(more formally, one verifies that, for any
integer valued $q>0$, 
$(\partial_\mu\varphi')^{2q}$ 
is a strongly irrelevant operator everywhere along
the spin wave critical line $K\geq0$). Note that this argument is nothing but
Harris criterion \cite{Harris 1974} in disguise.
Finally, by choosing 
$|{\bf y}_1-{\bf y}_2|$ sufficiently large, the two-point function 
$\langle 
\exp[{\rm i}\varphi({\bf y}_1)-{\rm i}\varphi({\bf y}_2)]
\rangle$
is seen to be a random variable with an arbitrarily small random component.
This is not so on all counts in the vortex sector.

\subsection{The vortex sector}
\label{The vortex sector}

Vortices in the $XY$ model are described by the field $\Theta$.
More precisely, the local density of vortices on the Euclidean plane
is given by $\partial^2_\mu\Theta$. 
Only the component $\tilde\partial_\mu\theta$ of the random vector potential
$A_\mu$ couples {\it directly} to the vortices described by $\Theta$.
Whereas the field $\Theta$ is induced by integer valued vortices,
the quenched disorder $\tilde\partial_\mu\theta$ 
describes real valued vortices.
Hence, the system tries to minimize the energy by screening 
the real valued quenched vortices with thermally excited integer valued 
vortices. However, by doing so, entropy is lost. The balance of energy 
and entropy could lead to a KT-like critical temperature 
separating a low temperature phase
with positive free energy and a high temperature phase with negative 
free energy. 

In fact in the absence of randomness in the spin stiffness,
the existence of a KT transition is suggested by a perturbative
RG calculation in the Coulomb (CB) gas representation  
\begin{eqnarray}
S^{\ }_{\rm CB}[\Theta,\theta]:&&=
E\sum_k (m_k-n_k)^2
\label{action for vortices}
\\
&&-
\pi K\sum_{k\neq l} 
(m_k-n_k)(m_l-n_l)\ln\left|{{\bf x}_k-{\bf x}_l\over l_0}\right|
\nonumber
\end{eqnarray}
of Eq. (\ref{factorized cal H}), provided $g_A$ is not too large
and {\it assuming} the existence of a dipole phase of the CB gas
at sufficiently low temperature and large reduced bare vortex core energy $E$
\cite{Rubinstein 1983,Fertig-Nattermann-Scheidl-Tang}.
Again, $\Theta$ is induced by a neutral configuration of vortices with 
vorticities $m_k\in{\bf Z}$, whereas $\theta$ is induced by a neutral
configuration of vortices with vorticities $n_l\in{\bf R}$.
It is sufficient to consider neutral configurations since the energy
cost of creating net vorticity scales logarithmically with the system size
$L$, $l_0$ being an arbitrary length scale. 

The perturbative RG analysis in the CB gas representation
is usually summarized by the phase diagram of Fig. 1. The phase diagram
is three dimensional with  $1/K=T/J$ the dimensionless temperature,
$g_A$ measuring the disorder strength,  
and $Y_1=\exp(-E)$ the fugacity of charge one vortex.
{\it All points on the plane with vanishing fugacity are critical}. 
This is the manifold of critical points describing the random spin wave phase.
Critical points within the shaded area are argued to be stable
\cite{Rubinstein 1983,Fertig-Nattermann-Scheidl-Tang}, 
i.e., $Y_1$ is irrelevant and thus decreases at long distances.
Critical points outside the shaded area are unstable, i.e.,
$Y_1$ is relevant and thus grows at long distances.

\begin{figure}
\begin{center}
\input{preprint8FIG11.pstex_t}
\hfill\break
\end{center}
FIG. 1.
\refstepcounter{figure}
\label{Fig. 1}
\small{
Proposed phase diagram for CB gas with quenched randomly 
fractionally charged vortices. 
$1/K$ is the reduced temperature, $g_A$ the variance
of the Gaussian disorder $\tilde\partial_\mu\theta$, 
$Y_1$ the charge one fugacity for thermal vortices.}\hfill
\end{figure}


To go beyond these results (still keeping $J$ non-random),
we prefer the Sine-Gordon (SG) representation of the CB gas 
in Eq. (\ref{action for vortices}) with $m_k=\pm1$ only. By an expansion
in powers of the ``magnetic field'' $h_1$ of the Boltzmann weight 
with Lagrangian
\begin{eqnarray}
{\cal L}^{\ }_{\rm SG}
[\chi,\theta]:=
{1\over2t}(\partial_\mu\chi)^2-{h_1\over t}\cos\chi
+{{\rm i}\over2\pi}\chi (\partial_\mu^2\theta)
\label{SG cal L},
\end{eqnarray}
followed by an integration over $\chi$,
we recover the grand canonical partition function of the charge one CB gas 
derived from Eq. (\ref{action for vortices}) provided one identifies 
\begin{equation}
K={t\over4\pi^2},\quad 
Y_1\sim {h_1\over2t},\quad 
\theta({\bf x})=\sum_{l=1}^Nn_l\ln\left|{{\bf x}-{\bf y}_l\over l_0}\right|.
\end{equation}

Notice that $\chi$ couples to the disorder $\tilde\partial_\mu\theta$ 
through a purely imaginary coupling, 
and that higher charges vortices $m_k=\pm2,\cdots$ are easily incorporated 
with higher harmonic $\cos(k\chi)$, $k\in{\bf N}$. Hence,
were it not for the ``magnetic field'' $h_1$, $\chi$ and $\theta$ would
decouple after the shift of integration variable 
\begin{equation}
\chi=:\chi'+{{\rm i}t\over2\pi}\theta
\end{equation}
very much in the same way the 
spin waves decouple from the longitudinal realizations of the disorder.

In the absence of disorder, one can establish the existence of the KT 
transition by performing a perturbative RG analysis on the two-point 
function
\begin{eqnarray}
\langle F_{{\bf x}_1,{\bf x}_2}\rangle^{\ }_0&&:=
{
\int{\cal D}[\chi]\ e^{-\int d^2{\bf x}{\cal L}_{\rm SG}[\chi,0]}\
e^{{\rm i}\chi({\bf x}_1)-{\rm i}\chi({\bf x}_2)}
\over
\int{\cal D}[\chi]\ e^{-\int d^2{\bf x}{\cal L}_{\rm SG}[\chi,0]}\
}
\nonumber\\
&&\equiv
{
\int{\cal D}[\chi]\ e^{-S_{\rm SG}[\chi,0]}\
e^{{\rm i}\chi({\bf x}_1)-{\rm i}\chi({\bf x}_2)}
\over
{\cal Z}_{\rm SG}[0]
}.
\label{pure case RG starting point}
\end{eqnarray}
In short, one first expands the right hand side of Eq.
(\ref{pure case RG starting point}) in powers of 
a very small fugacity $h_1/2t$. Without a short distance cutoff $a$,
all coefficients of the expansion in the fugacity are ill-defined.
The arbitrariness in the choice of the short distance cutoff is used
to derive RG equations obeyed by the fugacity and the reduced temperature.
The RG equations are integrated to determine whether the initial 
assumption of a very small fugacity is consistent. The irrelevance, 
marginality, and relevance of the fugacity then determines the spin wave phase,
KT transition, and disordered phase of the $XY$ model, respectively.

Rubinstein et al. \cite{Rubinstein 1983} followed the same strategy in the
presence of the quenched vector potential $\tilde\partial_\mu\theta$. 
More precisely, they performed a RG analysis
of the fugacity expansion of two correlation functions:
\begin{eqnarray}
&&
G_{{\bf x}_1,{\bf x}_2}:=-
\ln\langle F_{{\bf x}_1,{\bf x}_2}\rangle,
\label{RG of log of correlation function}
\\
&&
\langle F_{{\bf x}_1,{\bf x}_2}\rangle:=
{
\int{\cal D}[\chi]\ e^{-S_{\rm SG}[\chi,\theta]}\
e^{{\rm i}\chi({\bf x}_1)-{\rm i}\chi({\bf x}_2)}
\over
{\cal Z}_{\rm SG}[\theta]
},
\label{RG of correlation function}
\end{eqnarray}
to the first non-trivial order in the fugacity. Notice that
it is necessary to include
both $G_{{\bf x}_1,{\bf x}_2}$ and $\langle F_{{\bf x}_1,{\bf x}_2}\rangle$
to close the RG equations
to the first non-trivial order in the fugacity.
This is not surprising since taking the logarithm does not usually
commute with averaging.

The crucial point of the fugacity expansions in Eqs.
(\ref{RG of log of correlation function},\ref{RG of correlation function})
is that every expansion coefficients
depend on correlation functions calculated for
vanishing ``magnetic field'' $h_1$ (fugacity $h_1/2t$) such as
\begin{equation}
\int d^2{\bf y}_1\cdots d^2{\bf y}_{2n}
\overline{
\langle 
e^{
{\rm i}
[
\chi({\bf x}_1)-\chi({\bf x}_2)+
\chi({\bf y}_1)+\cdots-\chi({\bf y}_{2n})
]
}
\rangle_{h_1=0}
},
\end{equation}
on the one hand, but also such as
\begin{equation}
\overline{
\langle
e^{{\rm i}[\chi({\bf x}_1)-\chi({\bf x}_2)]}
\rangle_{h_1=0}
\left\langle
\int d^2{\bf y}_1d^2{\bf y}_{2}
e^{{\rm i}[\chi({\bf y}_1)-\chi({\bf y}_{2})]
}
\right\rangle^{n}_{h_1=0}
},
\end{equation}
on the other hand.
For vanishing fugacity $h_1/2t$, 
all averaged correlation functions are algebraic and
in particular [compare with Eq. (\ref{spin wave exponents})] 
\begin{eqnarray}
\overline{
\langle 
e^{{\rm i}\chi({\bf y}_1)}\ e^{-{\rm i}\chi({\bf y}_2)}
\rangle^q_{h_1=0}
}\ &&\propto\
{
\overline{
e^{-{qt\over2\pi}\theta({\bf y}_1)}\ e^{{qt\over2\pi}\theta({\bf y}_2)}}
\over|{\bf y}_1-{\bf y}_2|^{{qt\over2\pi}}}
\nonumber\\
\ &&=\
{1\over|{\bf y}_1-{\bf y}_2|^{2\pi Kq\left(1-g_AKq\right) } }.
\label{parabola}
\end{eqnarray}
This is our most important result. We will make use of it in 
section \ref{Non-analyticity of the fugacity expansion}
to deduce that the fugacity expansion cannot yield analytic 
scaling exponents around vanishing fugacity. 
The remarkable property of Eq. (\ref{parabola}) is that the scaling
exponent becomes negative for any given temperature $1/K$ and disorder 
strength $g_a$ as long as the moment $q$ is sufficiently large. Hence,
there must exist infinitely many local operators with negative scaling 
dimensions in the effective theory describing the random spin wave phase.

The correlation function in Eq. (\ref{parabola}) is closely related
to the strength of the interaction between
two external charges in the CB gas. Indeed, as we show in 
appendix \ref{Zero temperature considerations},
\begin{equation}
\langle 
e^{{\rm i}\chi({\bf y}_1)}\ e^{-{\rm i}\chi({\bf y}_2)}
\rangle_{h_1=0} = e^{-{K\over J}H_{1,2}}.
\label{exp H_12}
\end{equation}
Here, $H_{1,2}$ is the bare (since $h_1=0$) energy of two 
vortices of opposite unit charges 
in the background of the vector potential $\tilde\partial_\mu\theta$. 
The probability distribution of $H_{1,2}$ is calculated in
appendix \ref{Zero temperature considerations} and shown to be
a Gaussian distribution with a variance growing logarithmically with
$|{\bf y}_1-{\bf y}_2|$. Hence, the random variable
$\exp[(K/J)H_{1,2}]$ has a log-normal distribution. 
For a fixed separation  $|{\bf y}_1-{\bf y}_2|$,
the random energy $H_{1,2}$ can take arbitrarily 
negative values as a consequence of our initial assumption on
the probability distribution in Eq. (\ref{P[A_mu]}).  
This fact explains why the random variable
$\langle\exp\left[{\rm i}\chi({\bf y}_1)-{\rm i}\chi({\bf y}_2)\right]\rangle$
is unbounded from above.

Correspondingly, the ratio of the $q$ moment to the
first one raised to the power $q$
grows with $|{\bf y}_1-{\bf y}_2|$ raised to the positive power
$+2\pi g_AK^2q(q-1)$, in sharp contrast to the moments
of the logarithm of correlation functions 
[see Eq. (\ref{eq: variance vs mean of vortex energy})]
on the one hand, 
or to the moments of correlation functions in the spin wave sector
on the other hand.

We infer from Eq. (\ref{parabola}) that a sufficiently large moment 
of the two-point function in Eq. (\ref{parabola})
is not bounded from above
for arbitrary large values of the separation $|{\bf y}_1-{\bf y}_2|$.
This property is a consequence of $H_{1,2}$ being Gaussian distributed
that can also be understood as follows.
On the second line of Eq. (\ref{parabola}), the disorder average
is dominated by realizations of the disorder with $\theta({\bf y}_1)$
very negative and $\theta({\bf y}_2)$ very positive.
Such configurations are extremely rare for small separation 
$|{\bf y}_1-{\bf y}_2|$,
since the cost $(\tilde\partial_{\mu}\theta)^2$ will then be substantial, but
they become more likely as $|{\bf y}_1-{\bf y}_2|$ increases.

The sign of the $q^2$ dependency of the scaling exponents 
$2\pi Kq\left(1-g_AKq\right)$ on the right hand side of 
Eq. (\ref{parabola}) is thus the signature of broadly
distributed random variables. 
{\it We will show below that operators with the same scaling dimensions 
appear in the replicated action if random spin stiffness is introduced.}
The same scaling exponents have also been found by Korshunov to control
the fugacity expansion of correlation functions such as
$\overline{\langle F_{{\bf x}_1,{\bf x}_2}\rangle}$ to order $2q$ 
\cite{Korshunov 1993} 
(see section \ref{Non-analyticity of the fugacity expansion}
and appendix \ref{Fugacity expansion}).
Korshunov concluded from this property of the fugacity expansion
that any quasi-long-range ordered phase should be destroyed for 
arbitrary weak random vector potential $\tilde\partial_\mu\theta$.
An alternative conclusion, however, is that quasi-long-range order
is characterized by scaling exponents that are non-analytic functions
of the fugacity. We will come back to this scenario in section
\ref{Non-analyticity of the fugacity expansion}
and in the appendices.

The parabola 
\begin{equation}
g^{(1)}_A\left({1\over K}\right):=
{1\over K}\left(1-{2\over\pi}{1\over K}\right)
\label{g^(1)}
\end{equation}
(dashed line in Fig. 1) is obtained from the first moment $q=1$
by requiring that the scaling exponent on the right hand side of 
Eq. (\ref{parabola}) be ``marginal'', i.e., equals 4 \cite{Rubinstein 1983}. 
It is argued to delimit the boundary between quasi-long-range order 
and disordered phase for $\pi/4\leq 1/K\leq\pi/2$ \cite{Rubinstein 1983}. 

Instead of the reentrant phase transition for $0\leq1/K\leq\pi/4$,
the dilute vortex pair approximation proposed in
\cite{Fertig-Nattermann-Scheidl-Tang} 
suggests that the parabola should be replaced by the dotted line in Fig. 1. 
The difficulties with the fugacity expansion are here bypassed altogether 
since it is possible to calculate the mean of $G_{{\bf x}_1,{\bf x}_2}$
non-perturbatively in the fugacity provided it is assumed that:
$(i)$  {\it an insulating dipole phase exists} and
$(ii)$ {\it the interaction between dipoles can be neglected}.
Nevertheless, it remains an open problem to show rigorously that
at $T=0$ and for infinitesimal $g_A$, the ground state configuration
is in some quasi-long-range ordered phase consistent with assumptions
$(i)$ and $(ii)$ (see appendix \ref{Zero temperature considerations}).

\section{Non-analyticity of the fugacity expansion}
\label{Non-analyticity of the fugacity expansion}

We are now ready to describe the results obtained from the fugacity expansion
of correlation functions in the SG model Eq. (\ref{SG cal L}).
We will prove that the fugacity expansion is non-analytic.
For the pure system, non-analyticity of the fugacity expansion is interpreted
as the destruction of quasi-long-range order. 
In the presence of a random vector potential, we cannot rule out
the possibility that an exotic phase survives with 
quasi-long-range order characterized by scaling
exponents which are non-analytic functions of the fugacity.
However, even if a quasi-long-range ordered phase is present in the phase 
diagram of the SG partition function,
we will show in section \ref{The perturbative instability}
that randomness in the spin stiffness induces infinitely many
relevant perturbations to this critical behavior.

The mathematical reason for the breakdown of the fugacity expansion is
that we are expanding a random function in powers of a random variable
that takes values outside the radius of convergence of the expansion.
The physical reason for the breakdown is that the typical ground state
of the random phase $XY$ model does not support long-range-order 
(the ferromagnetic state). 
For the KT transition to survive the presence of a weak 
random vector potential $\tilde\partial_\mu\theta$,
the typical ground state must contain a sufficiently
large number of tightly bound pairs of vortices so as to destroy long-range
order, but not sufficiently large so as to screen the
bare logarithmic interactions of the vortices 
(see appendix \ref{Zero temperature considerations}).
Since the vortex fugacity measures, to a first approximation, the density of
vortices, a ground state with quasi-long-range order must imply the breakdown
of a fugacity expansion around vanishing fugacity 
(the ferromagnetic ground state).

Although the fugacity expansion is non-analytic, 
it is still useful to decide if non-analyticity 
reflects only that of scaling exponents of algebraic correlation functions 
or if it signals the breakdown of algebraic order.  
We have performed the fugacity expansion in the SG representation
to fourth order in the fugacity and could not, to this order, 
distinguish between an exotic algebraic phase 
from the complete destruction of quasi-long-range order. 

Lastly, the fugacity expansion is also instructive in that it allows to 
classify and understand the role played by
the large numbers of local operators that can be constructed within the
replica formalism. The close relationship between the fugacity expansion
and the replica formalism will be established below together with
appendix \ref{Fugacity expansion}
and section \ref{The perturbative instability}.

In principle, we would like to calculate the probability distribution
of the two-point functions  
$G_{{\bf x}_1,{\bf x}_2}$ and
$\langle F_{{\bf x}_1,{\bf x}_2}\rangle$.
This is done in appendix \ref{Zero temperature considerations} 
for vanishing fugacity. For finite fugacity, we are only able to calculate
their moments  perturbatively in the fugacity. 
We expect $G_{{\bf x}_1,{\bf x}_2}$ 
to be close to a Gaussian distribution since it is 
already Gaussian distributed for vanishing fugacity.
Hence, we will only calculate the mean of $G_{{\bf x}_1,{\bf x}_2}$.
On the other hand, we will need all moments of
$\langle F_{{\bf x}_1,{\bf x}_2}\rangle$
since it is log-normal distributed for vanishing fugacity.
Our goal is thus to calculate perturbatively in powers of the bare 
fugacity $h_1/2t$
\begin{eqnarray}
&&
\overline{G_{12}}:=\overline{G_{{\bf x}_1,{\bf x}_2}},
\label{mean CB interaction}
\\
&&
\overline{\langle F_{12}\rangle^{q}}:=
\overline{\langle F_{{\bf x}_1,{\bf x}_2}\rangle^{q}},\quad q\in{\bf N}.
\label{moments 2 point}
\end{eqnarray}

We restrict configurations of (thermal as well as quenched) vortices 
to neutral ones. This implies that only
even powers of the bare fugacity $h_1/2t$ enter in Eqs. 
(\ref{mean CB interaction},\ref{moments 2 point}).
The calculation to lowest order in the fugacity is summarized by
Eq. (\ref{parabola}). 

\subsection{Fugacity expansion up to second order}
\label{Fugacity expansion up to second order}

Up to second order in the bare fugacity $h_1/2t$,
we find (see appendix \ref{Fugacity expansion})
that the mean of the two-dimensional CB interaction between 
two external charges of opposite sign is
\begin{eqnarray}
\overline{G_{12}}\approx
2\pi
\left[\!
K-
4\pi^3 K^2 Y^2_{(1;1)}\!\!
\int\limits_1^{L/a}\!\!\! dy y^{3-2\pi\bar K}
\right]\!
\ln\left|{{\bf x}_{12}\over a}\right|,
\label{G_ 12 to second order in fugacity}
\end{eqnarray}
whereas
\begin{eqnarray}
&&
\overline{\left[\langle F_{12}\rangle\right]^q}\approx
\left|{{\bf x}_{12}\over a}\right|^{-2\pi x(q)},
\label{algebraic moment for F_12 to 2 order}
\end{eqnarray}
with the scaling exponents
\begin{eqnarray}
\label{F_ 12 to second order in fugacity}
x(q)&&:= 
\overline{K(q)} 
\label{x(q)}\\
&&- 
4\pi^3 q\left\{\left[{\overline{K(q)}\over q}\right]^2  
- K^4 g^2_A q^2 \right\}Y^2_{(1;1)}\!
\int\limits_1^{L/a}\!\! d y y^{3-2\pi\bar K}.
\nonumber
\end{eqnarray}
Here, we have introduced 
\begin{eqnarray}
&&
Y^2_{(1;1)} := \left({a^2h_1\over2t}\right)^2,
\\
&&
\bar K:= \overline{K(1)},\quad
\overline{K(q)}:= Kq -K^2 g^{\ }_Aq^2.
\label{overline K(q)}
\end{eqnarray}
The reason for which we label the 
dimensionless fugacity $Y_{(1;1)}$ by
the subindex $(1;1)$ will become clear to fourth order in the fugacity 
expansion. Suffices to say that to second order in the fugacity expansion, 
only one pair of vortices [of the type $(1;1)$] renormalizes the moments
of two-point functions.

The short distance (dimensionless) cutoff 
in the ubiquitous integral on the right hand sides of 
Eqs. (\ref{G_ 12 to second order in fugacity},\ref{x(q)}) is arbitrary. 
By splitting the range of integration into two ranges
$[1,e^l[$ and $[e^l,\infty[$ with $0<l\ll1$, it can be shown that
Eqs. 
(\ref{G_ 12 to second order in fugacity},
\ref{F_ 12 to second order in fugacity})
are form invariant provided $Y_{(1;1)}$ is renormalized multiplicatively,
and $K,\overline{K(q)}$ are renormalized additively. 
The disorder strength $g_A$ is left unchanged in this scheme.
We thus recover the well known RG equations
\cite{Rubinstein 1983}
\begin{eqnarray}
&&
{d Y^{\ }_{(1;1)}\over dl}=(2-\pi\bar K) Y^{\ }_{(1;1)},
\label{d h_1/ dl for q moment}
\\
&&
{d K\over dl}=-4\pi^3 K^2Y^2_{(1;1)},
\label{d K/ dl}
\\
&&
{d\overline{ K(q)}\over dl}=-
4\pi^3 q\left\{ 
\left[{\overline{K(q)}\over q}\right]^2 - 
K^4 g^2_A q^2 \right\}Y^2_{(1;1)},
\label{ d bar K(q)/ dl }
\\
&&
{d g^{\ }_A\over dl}=0.
\end{eqnarray}
So far so good, the scaling exponent $\bar K$
that controls the algebraic decay
of the two-point function $\overline{\langle F_{12}\rangle_{h_1=0}}$ gives
us a criterion for the breakdown of the fugacity expansion. 
All scaling exponents $x(q)$ are finite as long as
$4-2\pi\bar K\leq0$ up to second order in the bare fugacity $h_1/2t$.
This is not true anymore to fourth order in the bare fugacity.

Let us stress a few important points.
\begin{enumerate}
\item 
$Y_{(1,1)}$ appears in the RG equation of $K$. The latter controls the scaling
dimensions of many operators.
\item
When $4-2\pi \bar K < 0$, $Y_{(1,1)}$ flows to zero. The RG flow of $K$
[$\overline{K(q)}$] only changes $K$ [$\overline{K(q)}$]
by a finite amount proportional
to $Y_{(1;1)}^2$. Thus, scaling exponents only receive corrections of
order $Y_{(1;1)}^2$. We may then say that the scaling exponents are 
analytic in $Y_{(1;1)}$ around $Y_{(1;1)}=0$.
\item
When $4-2\pi \bar K > 0$, $Y_{(1,1)}$ flows to infinity at long distances
and $K$ blows up. This was interpreted as the instability of the
random spin wave critical point in Ref. \cite{Rubinstein 1983}. 
At the very least, scaling exponents are not 
analytic in $Y_{(1;1)}$ around $Y_{(1;1)}=0$ 
since the perturbative RG flow breaks down.
\end{enumerate}

\subsection{Fugacity expansion up to fourth and higher order}
\label{Fugacity expansion up to fourth and higher orders}

We have performed the fugacity expansion of 
Eqs. (\ref{mean CB interaction},\ref{moments 2 point})
to fourth order in the fugacity $h_1/2t$.
The fourth order calculation shows that, {\it due to the disorder}, 
the most singular coefficients in the fugacity expansions 
of $\overline{G_{12}}$ and $\overline{\langle F_{12}\rangle^q}$
are proportional to the integral
\begin{equation}
\lim_{{L\over a}\uparrow\infty}\int_1^{L\over a} dy y^{3-2\pi\overline{K(2)}}.
\label{integral defining boundary g(2)}
\end{equation}
Our derivation of the fugacity expansion of 
$\overline{\langle F_{12}\rangle}$ 
is sketched in appendix \ref{Fugacity expansion}.
It is also shown there that, in general, the most singular coefficient of
the fugacity expansion is proportional to the integral
\begin{equation}
\lim_{{L\over a}\uparrow\infty}\int_1^{L\over a} dy y^{3-2\pi\overline{K(n)}}
\label{most divergent integral to order 2n}
\end{equation}
to the order $2n$. Hence, to the $2n$-{th} order, the regime of analyticity
of the fugacity expansion is delimited by the curve 
[compare with Eq. (\ref{g^(1)})]:
\begin{equation}
{\rm min}
\left\{
g^{(1)}_A\left(1/K\right),\cdots,
g^{(n)}_A\left(1/K\right):=
{1\over Kn}\left(1-{2\over\pi}{1\over Kn}\right)
\right\}
\label{g^(n)}
\end{equation}
in the plane of vanishing fugacity in the three dimensional
coupling space depicted in Fig. 2. 

We must therefore conclude that the fugacity
expansions of Eqs. (\ref{mean CB interaction},\ref{moments 2 point})
are non-analytic 
for any infinitesimal value of the 
disorder strength $g_A$, since the region of analyticity in Fig. 2 shrinks 
to each new order and collapses to the segment $0\leq 1/K\leq\pi/2$, 
$g_A=0$ as $n\uparrow\infty$. Notice that
this argument breaks down in the pure system where
\begin{equation}
\lim_{g_A\downarrow0}\overline{K(n)}= Kn.
\end{equation}
This is nothing but the statement that $\cos(n\chi)$ 
is the more irrelevant the larger $n$ is.
To put it differently, it is necessary
that vortices be more irrelevant the 
higher their charges for the fugacity expansion to be analytic.

\begin{figure}
\begin{center}
\input{preprint8FIG21.pstex_t}
\hfill\break
\end{center}
FIG. 2.
\refstepcounter{figure}
\label{Fig. 2}
\small{
Boundaries $g^{(1)}_A(1/K)$ and $g^{(2)}_A(1/K)$ in the plane of vanishing
fugacity. The shaded area represents the regime of analyticity
of the fugacity expansion to fourth order in the fugacity.
}\hfill
\end{figure}


The integral in (\ref{integral defining boundary g(2)}) appears in the
RG flow to fourth order in the fugacity expansion. Indeed, we find that:
\begin{enumerate}
\item
\label{dbarK}
The RG equation of $\bar K$ contains a {\it new} fugacity $Y_{(1,1;2)}$.
\item
The RG equation for $Y_{(1,1;2)}$ is
\begin{equation}
\label{dY112}
{d Y^{\ }_{(1,1;2)}\over dl}=
\left[2-\pi\overline{ K(2)}\right] Y^{\ }_{(1,1;2)}.
\end{equation}
\end{enumerate}
Therefore, when $4-2\pi\overline{ K(2)} > 0$, $Y_{(1,1;2)}$ flows to infinity
and the RG flow of $\bar K$ blows up, suggesting the new phase boundary
$2\pi\overline{ K(2)}=4$, or, equivalently, $g^{(2)}_A(1/K)$. 
We believe that a similar structure appears to the order $2n$ in the fugacity
expansion:
\begin{enumerate}
\item
\label{dbarK (2n)}
The RG equation of $\bar K$ contains a {\it new} fugacity 
$Y_{(1,\cdots,1;n)}$.
\item
The RG equation for $Y_{(1,\cdots,1;n)}$ is
\begin{equation}
\label{dY1...1n}
{d Y^{\ }_{(1,\cdots,1;n)}\over dl}=
\left[2-\pi\overline{ K(n)}\right] Y^{\ }_{(1,\cdots,1;n)}.
\end{equation}
\end{enumerate}
The quasi-long-range phase boundary is controlled by
$2\pi\overline{ K(2n)}=4$, or, equivalently, $g^{(n)}_A(1/K)$. 
A stable quasi-long-range ordered phase can only
exist if $2\pi\overline{ K(2n)} > 4$ for all $n\in {\bf N}$. 
 
Based on the conventional RG analysis which essentially assumes that
one can switch off all fugacities but $Y_{(1,\cdots,1;n)}$ to order $2n$, 
we conclude that the quasi-long-range ordered phase is destroyed 
by any amount of disorder. A more conservative conclusion that we can draw
is that the perturbative RG flow must break down beyond some critical
order in the fugacity expansion that depends on the strength of the disorder.
In any case, {\it scaling exponents} cannot be analytic functions
of the bare vortex fugacity $Y_1$.

\section{The perturbative instability}
\label{The perturbative instability}

We have shown in section \ref{Broadly distributed correlation functions} 
that the critical theory describing the 
random spin wave phase must contain an infinite number of operators  with
negative scaling dimensions. It was shown in section 
\ref{Non-analyticity of the fugacity expansion} 
that these operators
with negative scaling dimensions have dramatic consequences on the 
RG equations within the fugacity expansion. We now complete the proof of the
existence of a perturbative instability by showing that the effective action
from which disorder averaged correlation functions are built is
necessarily perturbed by infinitely many operators with negative scaling 
dimensions.

The SG representation makes it clear that random spin stiffness 
cannot be dismissed as irrelevant as was the case in the spin wave sector
since it induces a random fugacity in addition to a random temperature.
Random spin stiffness has two very different consequences
from a symmetry point of view. 

To see this, we fermionize the SG model.
We use the (Euclidean) bosonization rules \cite{Coleman 1974}
\begin{eqnarray}
&&
\bar\psi{\rm i}\gamma_{\mu}\partial_{\mu}\psi\rightarrow
{1\over8\pi}(\partial_\mu\chi)^2,
\nonumber\\
&&
\bar\psi\gamma_\mu\psi\rightarrow
-{{\rm i}\over2\pi}\tilde\partial_\mu\chi,
\\
&&
-{\rm i}m_1\bar\psi\psi\rightarrow -{h_1\over t}\cos\chi,
\nonumber
\end{eqnarray}
to relate bilinears of Grassmann variables $\bar\psi,\psi$ 
to the real scalar field $\chi$ of Eq. (\ref{SG cal L}). 
Here, $\gamma_\mu$ are any two of the 
three Pauli matrices. Thus, the
grand canonical partition function of the CB gas 
derived from Eq. (\ref{action for vortices})
is equivalent to first expanding the 
partition function with Thirring Lagrangian
\begin{equation}
{\cal L}_{\rm Th}\!:=\!
\bar\psi({\rm i}\gamma_\mu\partial_\mu+{\rm i}m_1)\psi\!-\!
{g\over2}(\bar\psi\gamma_\mu\psi)^2\!+\!
(\bar\psi\gamma_\mu\psi)(\tilde\partial_\mu\theta)
\end{equation}
in powers of the mass ${\rm i}m_1$ and then integrating over
$\bar\psi,\psi$. The couplings $g,{\rm i}m_1$ are related to the reduced spin 
stiffness $K$ by
\begin{equation}
g:={1\over K}-\pi,\quad {\rm i}m_1\propto {h_1\over4\pi^2K}.
\end{equation}

Randomness in the vector potential of the $XY$ model 
thus enters the Thirring model through the coupling of the current
$j_\mu:=\bar\psi\gamma_\mu\psi$ 
of Dirac fermions to the transverse component $\tilde\partial_\mu\theta$
of the vector potential $A_\mu$. 
Randomness in the $XY$ spin stiffness enters the Thirring model through
a random mass and through a random current-current interaction. 
Moreover, the random mass ${\rm i}m_1$ and 
random Thirring interaction $g$ are generally not Gaussian distributed 
(even for a Gaussian distributed spin stiffness).

The fermionization of the random CB gas shows 
that random spin stiffness has two very different consequences
from a symmetry point of view.
On the one hand it induces a random mass which breaks the chiral symmetry
of the kinetic energy of the Dirac fermions. On the other hand it induces
a random current-current interaction which preserves the chiral symmetry.
Hence, any RG calculation should renormalize ${\rm i}m_1$ very differently
from $g$. 

In fact, it is sufficient to ignore randomness in $g$ altogether
for two reasons,
provided we include randomness in ${\rm i}m_1$ and $\tilde\partial_\mu\theta$.
Firstly, randomness in $g$ resembles 
non-Gaussian distributed vector potential $\tilde\partial_\mu\theta$.
It is then straightforward to show that non-Gaussian randomness in
$\tilde\partial_\mu\theta$ is irrelevant by calculating the scaling
dimension of $(\bar\psi\gamma_\mu\psi)^{2q}\sim(\partial_\mu\chi)^{2q}$
on the critical plane of Fig. 1.
Secondly, random ${\rm i}m_1$ is induced by a random fugacity of the CB gas
(random magnetic field of the SG model)
which will be seen to play a key role beyond the Villain approximation
of the random phase $XY$ model. Hence, 
it is meaningful to treat ${\rm i}m$, $g$ and $\tilde\partial_\mu\theta$, 
as independent random fields and to assume that only randomness in
${\rm i}m$ (or $h_1$ in SG model) and $\tilde\partial_\mu\theta$ are present.

To understand the interplay between a random fugacity $Y_1$
(magnetic field $h_1$) and
a random vector potential $\tilde\partial_\mu\theta$, we use
the replica formalism based on the identity
$\ln x=\lim_{r\downarrow0}(x^r-1)/r$. Indeed, within the replica approach,
disorder average can be performed directly 
on the replicated partition function.

Thus, starting from the replicated Lagrangian
\begin{eqnarray}
{\cal L}^{\ }_{\rm SG}&&=
\sum_{a=1}^r\left[
{1\over2t}(\partial_\mu\chi_a)^2-{h_1\over t}\cos\chi_a
+{{\rm i}\over2\pi}\chi_a (\partial_\mu^2\theta)
\right]
\nonumber\\
&&\!\sim
\sum_{a=1}^r\left[
\bar\psi_a({\rm i}\gamma_\mu\partial_\mu+{\rm i}m_1)\psi_a-
{g\over2}j_{a\mu}^2+\!
j_{a\mu}(\tilde\partial_\mu\theta)
\right],\!
\end{eqnarray}
one can integrate over the disorder in $h_1$ $({\rm i}m_1)$ or in 
$\tilde\partial_\mu\theta$.
In particular, integration over non-Gaussian $h_1$ $({\rm i}m_1)$
generates $r^q$, $q\in{\bf N}$, interactions through terms such as
\begin{equation}
\left[\sum_{a=1}^r (\cos\chi_a)({\bf x})\right]^q
\sim \left[\sum_{a=1}^r (\bar\psi_a\psi_a)({\bf x})\right]^q
\label{cumulant expansion if random core}
\end{equation}
in the replicated Lagrangian. According to conventional RG arguments,
the importance of these operators is measured by their scaling dimensions 
at a given critical point labeled by $1/K$ and $g_A$.
One then verifies along the derivation of Eq.
(\ref{parabola}) that, of all $r^q$ operators,  
\begin{equation}
{\cal O}_{1\cdots 1}\!:=\!\prod_{a=1}^{q  }\! e^{{\rm i}\chi_a}\!
\sim\!\prod_{a=1}^{q }\!\bar\psi_a\psi_a,\;
{\cal O}_{q}:=e^{{\rm i}q\chi_1}\sim(\bar\psi_1\psi_1)^q
\label{most and less relevant}
\end{equation}
are the most and less relevant, respectively, on the critical plane
$m_1=0$, $1/K\geq0$, $g_A\geq0$:
\begin{eqnarray}
&&
\overline{\langle 
{\cal O}^{\dag}_{1\cdots 1}({\bf x})\
{\cal O}^{\   }_{1\cdots 1}({\bf 0})\rangle}\propto
|{\bf x}|^{-2\pi Kq(1-g_AKq)},
\label{most relevant}
\\
&&
\overline{\langle 
{\cal O}^{\dag}_{q}({\bf x})\
{\cal O}^{\ }_{q}({\bf 0})\rangle}\propto
|{\bf x}|^{-2\pi Kq^2(1-g_AK)}.
\label{less relevant}
\end{eqnarray}
{\it The crucial point we want to make is that for any value of $g_A$, 
${\cal O}^{\ }_{1\cdots 1}$ becomes relevant
as $q$ is increased}. Hence, conventional RG arguments 
predict that all critical points on the plane 
$1/K\geq0$, $g_A>0$ are unstable with respect to random fugacity $Y_1$
(magnetic field $h_1$).

Next, we show that the existence of infinitely many relevant operators
is not an artifact of an ill-defined replica limit. Indeed,
there exists a very interesting line of critical points at twice the bare
KT transition temperature $1/K=\pi$, $g_A>0$, $Y_1=0$ 
in the phase diagram of Fig. 1.
{ }From the perspective of the Thirring model, it describes massless 
non-interacting Dirac fermions in the presence of a transverse random
vector potential. 
All disorder averaged correlation functions of local operators were
calculated with the help of supersymmetric techniques appropriate
to non-interacting systems \cite{Chamon196,Mudry96}. It was found 
in \cite{Chamon196,Mudry96} that each critical point has 
infinitely many primary fields and that these primary fields 
control the multifractal scaling of the random ``wave function'' 
$\exp[\theta({\bf x})]$. These primary fields are precisely given
by the family ${\cal O}_{1\cdots 1}$ in the replica language
and carry the scaling dimension $2q[1-(g_A/\pi)q]$ appropriate to $1/K=\pi$.

If we accept the hypothesis that having infinitely
(as opposed to finitely) many relevant perturbations results in
the instability of the line $1/K=\pi$, $g_A>0$, we need not even rely
on the replica approach to rule out a KT transition.
Indeed, the shape of the phase digram of Fig. 1 is preserved
if we forbid thermal excitations of vorticity one ($Y_1=0$)
but allow thermal excitations of vorticity two ($Y_2\geq0$).
More precisely, if we replace in the SG (Th) Lagrangian 
${h_1\over t}\cos\chi$ (${\rm i}m_1\bar\psi\psi$) 
by ${h_2\over t}\cos(2\chi)$ (${\rm i}m_2(\bar\psi\psi)^2$),
then the parabola
$g_A({1\over K})={1\over K}(1-{2\over\pi}{1\over K})$
is simply rescaled to
${1\over K}(1-{1\over2^2}\times{2\over\pi}{1\over K})$.
The inescapable conclusion is then that random fugacity for charge 
two vortices destroys
this quasi-long-range ordered phase since it generates relevant interactions
given by Eq. (\ref{most relevant}) under length rescaling.

\section{The Villain approximation}
\label{The Villain approximation}

In this section, we go back to the lattice to study in more details the nature 
of the Gaussian approximation made in Eq. (\ref{eq: H_G}). 
We consider both the random bond $XY$ and Villain models on the square lattice.
We will see that the difficulties with the fugacity expansion are not associated
with pathologies of the field theory at short distances (such as ill-defined
operator product expansions) but are intrinsic to 
the fugacity expansion, i.e., are also present if the fugacity expansion is
performed on the Villain model itself. We will also see that randomness
in the phase only ($g_A>0$, $g_J=0$)
has different consequences in the $XY$ and Villain models.

We begin with the random partition function
\begin{mathletters}
\begin{eqnarray}
{\cal Z}^{\ }_{XY}&&:=
\int\limits_0^{2\pi}\!\!
\left(\prod_{i=1}^{L^2}{d\phi_i\over2\pi}\right)
\prod^{2L^2}_{\langle ij \rangle}
e^{-{\cal L}^{XY}_{ij}},
\label{XY partition func}\\
{\cal L}^{XY}_{ij}&&:=
K_{ij}\left[1-\cos(\phi_i-\phi_j-A_{ij})\right],
\label{XY link energy}
\end{eqnarray}
\end{mathletters}
\noindent
for the random bond $XY$ model on a square lattice made of $L^2$ sites.
Directed links (two per site) on the square lattice
are denoted by $\langle ij \rangle$.
The phases $A_{ij}=-A_{ji}$ 
are random (with short-range correlations for different links). 
However, they need not be restricted to
$0\leq A_{ij}<2\pi$  in spite of the periodicity of the cosine. Indeed,
the probability distribution for $A_{ij}$ need not be periodic with period
$2\pi$. The reduced spin stiffness $J_{ij}>0$ are also random 
(with short-range correlations for different links).
A reasonable choice for the probability distribution of the
random phases in the $XY$ model is
\begin{equation}
P[A^{\ }_{ij}]:=
\prod_{\langle ij\rangle}^{2L^2}
{1\over\sqrt{2\pi g_A}}\ e^{-{1\over2g_A}A^2_{ij}}.
\label{P[A_ ij]}
\end{equation}
This choice is made without loss of generality provided any ``small'' departure
from Eq. (\ref{P[A_ ij]}) does not prevent the system to flow to 
the fixed point it would have reached otherwise.
Although the random phase can take all possible real values, the
energy per link ${\cal L}^{XY}_{ij}$ cannot take values outside the range
(assumed compact for any {\it finite} temperature)
\begin{equation}
0\leq {\cal L}^{XY}_{ij}\leq 
2\sup_{\langle ij\rangle}\ K_{ij}
\end{equation}
with probability one.

The random Villain model consists in defining on the same lattice 
the random partition function 
\cite{Villain 1975,Jose 1977,Fradkin 1978,Paczuski 1991}
\begin{eqnarray}
{\cal Z}_{\rm V}&&:=
e^{-KL^2}
\int\limits_0^{2\pi}
\left(\prod_{i=1}^{L^2} {d\phi_i\over2\pi}\right)
\prod^{2L^2}_{\langle ij \rangle}\sum_{l_{ij}\in{\bf Z}}
e^{-{\cal Q}^{\rm V}_{ij}},
\nonumber\\
{\cal Q}^{\rm V}_{ij}&&:=
{K_{ij}\over 2}\left(\phi_i-\phi_j-A_{ij}-2\pi l_{ij}\right)^2.
\label{Villain partition func}
\end{eqnarray}
Here, we have taken the spin stiffness to be selfaveraging, i.e.,
\begin{equation}
K:={1\over2L^2}\sum_{\langle ij\rangle}^{2L^2}K_{ij}=
{J\over T}+{\cal O}\left({1\over L}\right).
\end{equation}
The periodicity under a shift of any $\phi_i$ by $2\pi$ is preserved in
the Villain action, but the non-linearity of the cosine has been removed in
the Villain action. When referring to the random Villain model, 
we will have in mind
the {\it same} probability distribution for the spin stiffness and for 
the random phase as for the $XY$ model.

The quantity ${\cal Q}^V_{ij}$ is not the counterpart to the link energy
(\ref{XY link energy}) 
in the $XY$ model since it is not periodic under a shift of $A_{ij}$ 
or $\phi_i$ by
an integer multiple of $2\pi$. It is, however, very closely related to the
energy of a given
configuration of vortices in the background of a random environment
induced by bond randomness.

By taking the random phase $A_{ij}$ of the Villain model to be Gaussian 
distributed according to Eq. (\ref{P[A_ ij]}), we immediately see that
${\cal Q}^{\rm V}_{ij}$ can take any arbitrary large value with
a finite probability. Moreover, $\exp\, ({\cal Q}^{\rm V}_{ij})$ is
log-normal distributed very much in the same way as correlation functions
for vortex operators are in the Gaussian approximation
(see section \ref{Broadly distributed correlation functions} 
and appendix \ref{Zero temperature considerations}).
This behavior should be contrasted with that of the Villain link energy
\begin{equation}
{\cal L}^{\rm V}_{ij}:=-
\ln\left(\sum_{l_{ij}\in{\bf Z}}e^{-{\cal Q}^{\rm V}_{ij}}\right),
\end{equation}
which is indeed periodic
under a change of $A_{ij}$ or $\phi_i$
by an integer multiple of $2\pi$.
It is crucial to realize that periodicity of the Villain link energy
${\cal L}^{\rm V}_{ij}$ is broken to any finite order in a fugacity expansion,
since the fugacity expansion amounts to a truncation of the summation over
$l_{ij}\in{\bf Z}$.

There is a noteworthy difference between the random bond
$XY$ and Villain models.
If we take the spin stiffness to be non-random but allow the
relative phase between neighboring spins to be random, then
we must assume that the SG model has both a random vector potential
and a random fugacity if it is interpreted as the minimal model capturing
the long distance properties of the $XY$ model. To the contrary,
the fugacity of the SG model is not random if it is derived from the Villain
model with random phase but no randomness in the spin stiffness.


To clarify these last two points, we need first to introduce some notation. 
We define the longitudinal and transversal components
$A^{\|  }_{ij}$ and $A^{\bot}_{ij}$, respectively, by
\begin{eqnarray}
A^{\ }_{ij}&&:= A^{\|  }_{ij} + A^{\bot}_{ij},
\end{eqnarray}
where $A^{\|  }_{ij}$ is curl free, i.e.,
\begin{eqnarray}
0&&={\rm curl}_{\bf i}\ A^{\|  }_{ij}
\\
&&:=\!
A^{\|  }_{i(i+\hat{\bf x})}\!+\!
A^{\|  }_{(i+\hat{\bf x})(i+{\bf x}+\hat{\bf y})}\!+\!
A^{\|  }_{(i+\hat{\bf x}+{\bf y})(i+\hat{\bf y})}\!+\!
A^{\|  }_{(i+\hat{\bf y})i},
\nonumber
\end{eqnarray}
and $A^{\bot}_{ij}$ is divergence free, i.e.,
\begin{eqnarray}
0&&={\rm div}_{i}\  A^{\bot}_{ij}
\nonumber
\\
&&:= 
A^{\bot}_{i(i+\hat{\bf x})}-A^{\bot}_{(i-\hat{\bf x})i}+
A^{\bot}_{i(i+\hat{\bf y})}-A^{\bot}_{(i-\hat{\bf y})i}.
\end{eqnarray}
Dual sites are labeled by
\begin{eqnarray}
{\bf i}&&:= i+{1\over2}\hat{\bf x}+{1\over2}\hat{\bf y},
\end{eqnarray}
where the basis vectors $\hat{\bf x}$ and $\hat{\bf y}$
span the square lattice.
For completeness, the gradient of a lattice scalar is defined by
\begin{equation}
{\rm grad}_{\hat\mu} \phi_i:= \phi_{i+\hat\mu}-\phi_{i},
\qquad \hat\mu=\hat{\bf x},\hat{\bf y}.
\end{equation}

It is possible to rewrite the Villain partition function solely in terms
of degrees of freedom defined on the dual lattice 
\cite{Villain 1975,Jose 1977,Fradkin 1978,Paczuski 1991},
\begin{eqnarray}
{\cal Z}_{\rm V}&&\propto\!\!
\sum_{\{l^{\bot}_{\{ {\bf i}{\bf j}} \}\in{\bf Z}^{2L^2}}
\int\limits_{-\infty}^{+\infty}
\left(\prod_{{\bf i}=1}^{L^2} 
{d\varphi_{{\bf i}}\over2\pi}\right)
e^{
-{1\over 2}\sum\limits^{2L^2}_{\langle{\bf i}{\bf j}\rangle}
\left[
{\cal Q}^{\rm SW}_{{\bf i}({\bf i}+\hat\mu)}+
{\cal Q}^{\rm CB}_{{\bf i}({\bf i}+\hat\mu)}
\right]
}.
\nonumber\\
\end{eqnarray}
We have dropped a multiplicative factor that depends on the 
(random) spin stiffness, and
\begin{eqnarray}
&&
{\cal Q}^{\rm SW}_{{\bf i}({\bf i}+\hat\mu)}+
{\cal Q}^{\rm CB}_{{\bf i}({\bf i}+\hat\mu)}:=
\label{dual Villain action}
\\
&&
{
\left({\rm grad}_{\hat\mu}\varphi_{{\bf i}}\right)^2
\over K^{\ }_{{\bf i}({\bf i}+\hat\mu)}}+
2{\rm i}\left({\rm grad}_{\hat\mu}\varphi_{{\bf i}}\right)
\left(A^{\bot}_{{\bf i}({\bf i}+\hat\mu)}+
2\pi l^{\bot}_{{\bf i}({\bf i}+\hat\mu)}\right).
\nonumber
\end{eqnarray}
The random dual phases 
$A^{\bot}_{{\bf i}{\bf j}}$ are now curl free as are
dual vortex degrees of freedom $l^{\bot}_{{\bf i}{\bf j}}$.

Since Eq. (\ref{dual Villain action}) 
is quadratic in the spin wave degrees of freedom $\varphi_{{\bf i}}$
whereas Eq. (\ref{dual Villain action})  
is linear in the vortex degrees of freedom
$l^{\bot}_{{\bf i}{\bf j}}$, 
it is possible to decouple the spin wave sector from the vortex sector. 
Our final expression for the partition function in the vortex sector is
[compare with Eq. (\ref{action for vortices})]
\begin{eqnarray}
&&
{\cal Z}_{\rm CB}=
\sum_{\{m_{{\bf i}}\}}
\prod^{L^2}_{{\bf i},{\bf j}=1} 
e^{-\pi 
(m_{{\bf i}}-n_{{\bf i}})(m_{{\bf j}}-n_{{\bf j}})
D_{{\bf i}{\bf j}}[K_{{\bf k}{\bf l}}]},
\\
&&
{\rm div}_{{\bf i}} l^{\bot}_{{\bf i}{\bf j}}=: 
{1\over\sqrt{2\pi}} m_{{\bf i}},
\\
&&
{\rm div}_{{\bf i}}A^{\bot}_{{\bf i}{\bf j}}=: 
\sqrt{2\pi} n_{{\bf i}}.
\end{eqnarray}
Here, $D_{{\bf i}{\bf j}}[K_{{\bf k}{\bf l}}]$ 
are the random components of the dual lattice Green function 
in the background of random spin stiffness,
namely the inverse of the quadratic form in the spin wave sector.

Performing a dual transformation on the random bond Villain model thus offers
two insights. First, given a Gaussian probability distribution 
for the random phases and no randomness in the spin stiffness, 
the probability distribution for the CB energy
of a given configuration of vortices $\{m_{\bf i}\}$ 
is Gaussian, and exponentiating this CB energy
yields a log-normal distributed random variable. We thus conclude that 
{\it the existence of log-normal
distributed correlation functions in the random spin wave phase is not
an artifact of the continuum limit}.

Second, the diagonal components $D_{{\bf i}{\bf i}}[K_{{\bf k}{\bf l}}]$
of the Green function define the vortex core energy. 
{\it Hence, in the absence
of random spin stiffness, the core energy of the vortex sector is not random
within the Villain model}.

We recover the $XY$ partition function by replacing the right hand side of
Eq. (\ref{Villain partition func}) by
\begin{equation}
-K_{ij}\sum_{n=1}^\infty{(-1)^n\over(2n)!}
\left(\phi_i-\phi_j-A_{ij}-2\pi l_{ij}\right)^{2n}.
\label{non-linearities}
\end{equation}
Spin waves and vortices are coupled in the $XY$ model with or without 
randomness. In an effective theory such as the SG model, non-linearities
such as in Eq. (\ref{non-linearities}) 
can be incorporated through a random fugacity to a first approximation. 
In the clean limit, the irrelevance of higher vorticity charges
justifies neglecting such an effect. However, as we have shown in section 
\ref{The perturbative instability}
this is not anymore the case in the presence of random phases.

A random core energy for vortices is always induced by a random spin stiffness.
However, we have shown that the $XY$ and Villain models
differ in one very important aspect when only the relative phase between 
neighboring spins is random. Indeed, due to non-linear effects the core energy
of vortices is always random in the $XY$ model with random phase
whereas this is not the case for the Villain model. If randomness 
in the core energy does indeed destroy the exotic quasi-long-range 
order proposed in \cite{Fertig-Nattermann-Scheidl-Tang},
we must then conclude that the Villain model with random phase only
does not belong to the same universality class as the $XY$ model 
with random phase only {\it in a strong sense}. It would be very interesting 
to probe numerically our conjectured difference between 
the random $XY$ and Villain models. 

We have also shown that the random $XY$ model and any approximative treatment
of the Villain model based on a perturbative expansion in the fugacity
do not belong to the same universality class in {\it a weak sense}.
We can illustrate this fact in a very suggestive way. We take
the spin-spin correlation function (thermal average)
in the random bond $XY$ model for an arbitrary pair of sites $i$ and $j$. 
Clearly, this is a random variable that takes values on a compact range,
namely
\begin{equation}
|\langle\cos \phi_i \cos \phi_j \rangle|\leq 1.
\end{equation}
Consequently, all integer moments with $p<q$ obey
\begin{eqnarray}
&&
\overline{|\langle\cos \phi_i \cos \phi_j \rangle|^p}\leq 1,
\label{bound one}
\\
&&
\overline{|\langle\cos \phi_i \cos \phi_j \rangle|^q}\leq 
\overline{|\langle\cos \phi_i \cos \phi_j \rangle|^p}\leq1. 
\label{bound two}
\end{eqnarray}
It is possible to verify that those inequalities 
are not satisfied to any finite order in a fugacity expansion of the
random phase Villain model. Without loss of generality, we can take the
continuum limit. We recall that the prescription to estimate
the spin-spin correlation function (thermal average)
from the continuum theory of section
\ref{Broadly distributed correlation functions} is to identify 
\cite{Jose 1977}
\begin{equation}
\langle\cos \phi_i \cos \phi_j \rangle^{\ }_{{\cal Z}_{XY}}
\rightarrow
\left\langle 
e^{-{1\over 2\pi K}\int^{{\bf x}_i}_{{\bf x}_j}d s_\mu\tilde\partial_\mu\chi}
\right\rangle^{\ }_{{\cal Z}_{\rm SG}}
\end{equation}
for some path joining ${\bf x}_j$ to ${\bf x}_i$. To any finite order in a
fugacity expansion, the spin-spin correlation function (thermal average)
is very broadly distributed
(log-normal in the limit of infinite core energy), and thus violates the 
bounds of Eqs. (\ref{bound one},\ref{bound two}). We stress that this is
a failure of the fugacity expansion on the
Villain model itself and not an artifact of the continuum limit
\footnote
{
An alternative way to stress this point is to note that the
exact identity between the two- and four-points spin-spin
correlation functions \cite{Ozeki 1993}
\begin{equation}
\overline{\langle e^{{\rm i}(\phi_i-\phi_j)}\rangle}=
\overline{|\langle e^{{\rm i}(\phi_i-\phi_j)}\rangle|^2}
\end{equation}
that holds along the Nishimori line $K=1/g^{\ }_A$ in the Villain model
is always violated to any finite order in the fugacity expansion.
}.

This situation is very reminiscent of that in the 
one-dimensional random bond Ising model.
For a large but fixed separation, 
the Ising spin-spin correlation function (thermal average) is ``close'' 
to being log-normal \cite{Kardar 1996}. The log-normal approximation
describes exactly the mean and variance of the logarithm of the
spin-spin correlation function (thermal average)
in the Ising case but it neglects higher cumulants. 
By analogy, we might expect
that the fugacity expansion on the Villain model
captures well the logarithm of the $XY$ spin-spin correlation (thermal average).
However, the log-normal approximation in the random bond Ising model
fails badly to describe the tails of the Ising spin-spin correlation functions
very much in the same way as the fugacity expansion on the Villain model
dramatically overestimate tails for the probability
distribution of the $XY$ spin-spin correlation functions.

So it is by now clear that the probability distribution
of the random vector potential unduly favors rare events through its tails
within the fugacity expansion. In one scenario, 
we must then anticipate partial loss of universality
for the fixed point probability distribution of correlation
functions (thermal average)
that are broadly distributed, since their tails result from rare events. 
We are aware of several examples of this kind:
directed polymers in a random medium \cite{Kardar 1996}, 
the metal-insulator transition \cite{AKL 1991,Falko 1995,Smolyarenko 1997},
and quantum gravity \cite{Kogan96}. 
In the worst case scenario, the fugacity expansion on the
Villain model is overwhelmed by rare events and loses complete
predictability with regard to the phase diagram of the $XY$ model.

\section{Conclusions}
\label{Summary}

Concerned with the possibility
that effective theories describing random critical points are often
characterized by a spectrum of scaling exponents that is unbounded
from below, we have studied the Gaussian approximation to the random $XY$ 
model on a square lattice within a perturbative RG framework.

The Gaussian approximation to the random $XY$ model predicts the existence
of a manifold of random critical points describing a random spin wave phase.
However, there are correlation functions in the random spin wave phase that are
log-normal distributed. Correspondingly, there are infinitely many operators
with negative scaling dimensions that are associated with vortices in the 
effective theory describing the random spin wave phase. 

We showed that all these operators with negative scaling dimensions contribute
in a highly non-trivial way to the perturbative RG equations within a fugacity
expansion around the random spin wave phase. The existence of
infinitely many negative scaling dimensions thus manifests itself by the
non-analyticity of the fugacity expansion for any given temperature and 
disorder strength. In this sense, the random spin wave phase is unstable,
although we cannot preclude the possibility that a new phase with
quasi-long-range order can be found for sufficiently low temperatures and
weak disorder strength. The breakdown of the fugacity expansion is also 
associated to a potential perturbative instability triggered by 
any random vortex core energy.

We have shown that neither the breakdown of the fugacity expansion nor the
perturbative instability are artifacts of the continuum approximation 
that we used but would also be present in the random bond Villain model on 
the square lattice. Rather, they both reflect the extreme sensitivity of the 
fugacity expansion to the tails of the probability distribution 
that is chosen for the random bonds.

The physical interpretation for the breakdown of the fugacity expansion is that 
the typical ground state is not ferromagnetic. 
In the best case scenario, 
the typical ground state for weak disorder supports some quasi-long-range order that would persist
for sufficiently low temperatures. 
However, to address the nature of the low temperature,
weak disorder region of the phase diagram it is imperative to better 
characterize the typical ground state and to use a RG scheme that is 
non-perturbative in the vortex fugacity.

\acknowledgments

We are grateful to L. Balents, P.W. Brouwer, J.L. Cardy, 
J. Chalker, E. Fradkin, M. Kardar, I. Lerner, and A.W.W. Ludwig, 
for stimulating discussions.
This work was completed in parts while we
were participants at the Program on 
{\it Quantum Field Theory in Low Dimensions} 
at the Institute for Theoretical Physics of the 
University of California at Santa Barbara. 
We are very grateful to the organizers of the program and particularly 
to Prof. Jim Hartle, Director of ITP, for his warm hospitality.
This work was supported in part by the NSF through the grants 
NSF DMR97-14198  
at the Massachusetts Institute of Technology, NSF PHY94-07194 
at the Institute for Theoretical Physics of the University of California at 
Santa Barbara, and by (CM) the Swiss Nationalfonds.

\appendix

\section{Zero temperature considerations}
\label{Zero temperature considerations}

In this appendix, we study some properties of the random phase $XY$
model at zero temperature. Indeed, we recall that the prerequisite 
to the existence of a KT transition in the disorder free $XY$ model
is the simple fact that the ground state of the Hamiltonian
\begin{equation}
H[\{\phi_i\}]:=
\sum_{\langle ij \rangle} J[1-\cos(\phi_i-\phi_j)]
\end{equation}
is ferromagnetic. In two dimensions,
the long-range order of the ground state is first
demoted to quasi-long-range order for any finite temperature
below the KT transition temperature $T_{\rm KT}$. In turn,
topological excitations in the form of vortices
wipe out the quasi-long-range order for $T>T_{\rm KT}$.
The crucial question we would like to address is:
are typical ground states of 
\begin{equation}
H[\{\phi_i\};\{A_{ij}\}]:=
\sum_{\langle ij \rangle} J [1-\cos(\phi_i-\phi_j-A_{ij})]
\end{equation}
ordered, quasi-long-range ordered, or disordered?

What we mean by a typical ground state is the following.
We assume that we know how to calculate the probability 
to find a ground state with energy density $e$  
(energy divided by total number of sites, i.e., $0\leq e \leq 4J$).
A typical ground state maximalizes this probability distribution.
Unfortunately, obtaining this probability distribution is very difficult
in view of: $(i)$ the non-quadratic dependency of the energy on the disorder,
and $(ii)$ of the need to minimalize the energy spectrum of 
$H[\{\phi_i\};\{A_{ij}\}]$ for a given realization of the disorder. 
The first difficulty can be disposed of in the Villain approximation
(whereby it is assumed that the Villain approximation does not change the
universality class), but we still must face the second difficulty.

We focus on the vortex sector in the continuum approximation of section 
\ref{Broadly distributed correlation functions} and show that, 
for extensively many realizations of the random vector potential 
$\tilde\partial_\mu\theta$, the ground states are, 
for lack of a better description, {\it complex} configurations of vortices. 
We do this by generalizing an argument used in 
\cite{Fertig-Nattermann-Scheidl-Tang} 
to prove that the random potential $\tilde\partial_\mu\theta$
destroys the long-range order of the pure system at $T=0$ and  
for sufficiently strong disorder strength: $g_A>\pi/8$.
We then modify the random phase $XY$ model by restricting
the possible realizations of the disorder to a more tractable subset.
Within this subset we show that it is possible to decide whether
the ground states support quasi-long-range order or not.

\subsection{Gaussian distribution of the energy of vortex configurations}
\label{Gaussian distribution of the energy of vortex configurations}

Let $\tilde\Theta$ in Eq. (\ref{factorized cal H}) 
be given by the vortex configuration
\begin{equation}
\tilde \Theta({\bf x}):=
\sum_{i=1}^n m_i\ln\left|{{\bf x}-{\bf x}_i\over l_0}\right|,
\quad m_i\in{\bf Z},
\end{equation}
(which need not be neutral)
and define its energy 
in the background of the random vector potential 
$\tilde\partial_\mu\theta$
to be 
\begin{eqnarray}
H_{1,\cdots,n}&&:=
{J\over2}\int d^2{\bf x}\ 
\left[
(\tilde\partial_\mu\tilde\Theta)^2-
2(\tilde\partial_\mu\tilde\Theta)(\tilde\partial_\mu\theta)
\right]
\nonumber\\
&&\equiv\
\overline{H_{1,\cdots,n}}+
\delta H_{1,\cdots,n},
\label{H{1...n}}
\end{eqnarray}
where
\begin{eqnarray}
&&
\overline{H_{1,\cdots,n}}=
-\pi J
\sum_{k,l=1}^n m_k m_l \ln\left|{{\bf x}_k-{\bf x}_l\over l_0}\right|,
\nonumber\\
&&
\delta H_{1,\cdots,n}=
2\pi J\!\sum_{k=1}^n\! m_k\theta({\bf x}_k).
\label{ H 1 to n}
\end{eqnarray}
The definition in Eq. (\ref{H{1...n}}) is useful since the energy
$H_{1,\cdots,n}$ is very closely related to the 
thermal average of the $2m=n$-point correlation function
\begin{equation}
F_{{\bf x}_1,\cdots,{\bf x}_{2m}}:= 
e^{ {\rm i}\chi({\bf x}_1)    }\cdots 
e^{ {\rm i}\chi({\bf x}_m)    }
e^{-{\rm i}\chi({\bf x}_{m+1})}\cdots 
e^{-{\rm i}\chi({\bf x}_{2m })},
\label{F(1 to 2m)}
\end{equation}
for vanishing fugacity $h_1/2t$. Indeed, in that case
\begin{equation}
\langle F_{{\bf x}_1,\cdots,{\bf x}_{2m}}\rangle_{h_1=0}= 
e^{-{K\over J} H_{1,\cdots,2m}}.
\end{equation}

Notice that for a {\it fixed} realization of $\tilde\partial_\mu\theta$, 
$H_{1,\cdots,n}$ is {\it bounded from below} by 
$-{J\over2}\int d^2{\bf x}(\tilde\partial_\mu\theta)^2$. On the other hand,
since
${J\over2}\int d^2{\bf x}(\tilde\partial_\mu\theta)^2$
can take arbitrary large values, these are the rare events which take full
advantage of the non-compactness of the Gaussian probability distribution
for $\tilde\partial_\mu\theta$,
the  probability distribution $P(E;{\bf x}_1,\cdots,{\bf x}_n)$
that $H_{1,\cdots,n}$ takes the value $E$ is non-vanishing 
for all real values of $E$. A very special vortex configuration
is the {\it ferromagnetic} configuration $\tilde\Theta=0$. This is
the ground state of the pure system and it has a vanishing random energy.
Clearly it need not be the ground state for a given realization of
the disorder.

What is the probability distribution of $H_{1,\cdots,n}$?
By definition it is given by
\begin{eqnarray}
&&
P(E;{\bf x}_1,\cdots,{\bf x}_n):=
\nonumber\\
&&
{
\int{\cal D}[\theta]\ 
e^{-{1\over2g_A}\int d^2{\bf y}\ (\partial_\mu\theta)^2({\bf y})}\
\delta(E-H_{1,\cdots,n})
\over
\int{\cal D}[\theta]\ 
e^{-{1\over2g_A}\int d^2{\bf z}\ (\partial_\mu\theta)^2({\bf z})}
}.
\label{def P(E; x1...xn)}
\end{eqnarray}
We can represent the delta function by an integral, in which case
\begin{eqnarray}
P(E;{\bf x}_1,\cdots,{\bf x}_n)
&&=
\int{d\lambda\over2\pi}\ 
e^{{\rm i}\lambda\left(E-\overline{H_{1,\cdots,n}}\right)}\
\overline{e^{-{\rm i}\lambda\delta H_{1,\cdots,n}}}
\nonumber\\
&&=
\int{d\lambda\over2\pi}\ 
e^{{\rm i}\lambda\left(E-\overline{H_{1,\cdots,n}}\right)}\
e^{
-{\lambda^2\over2}
\overline{\left(\delta H_{1,\cdots,n}\right)^2}
}
\nonumber\\
&&=
{
\exp\left[{-{\left(E-\overline{H_{1,\cdots,n}}\right)^2\over2
\overline{\left(\delta H_{1,\cdots,n}\right)^2}}}\right]
\over\sqrt{2\pi 
\overline{\left(\delta H_{1,\cdots,n}\right)^2}}}.
\label{formula for P(E; x1...xn)}
\end{eqnarray}
Thus, the probability distribution for the energy
of the vortex configuration 
$\tilde\Theta({\bf x})$
is Gaussian with mean
$\overline{H_{1,\cdots,n}}$
and variance 
\begin{eqnarray}
\overline{\left(\delta H_{1,\cdots,n}\right)^2}&&=
-2\pi J^2g_A\sum_{k,l=1}^n 
m_k m_l\!\ln\left|{{\bf x}_k-{\bf x}_l\over l_0}\right| 
\nonumber\\
&&=
2J g_A\ \overline{H_{1,\cdots,n}}.
\label{eq: variance vs mean of vortex energy}
\end{eqnarray}

We are now ready to calculate the (unnormalized) distribution
$P(E;m_1,\cdots,m_n)$ to find $n$ vortices with vorticities
$m_1,\cdots,m_n\in{\bf Z}$ 
anywhere on the Euclidean plane. It is defined by
\begin{eqnarray}
&&
P(E;m_1,\cdots,m_n):=
\nonumber\\
&&
\int {d^2{\bf x}_1\over a^2}\cdots {d^2{\bf x}_n\over a^2}
P(E;{\bf x}_1,\cdots,{\bf x}_n),
\end{eqnarray}
where $a$ is a microscopic cutoff and $L$ is the system size.
We are interested in the scaling of 
$P(E;m_1,\cdots,m_n)$
with the system size $L$.
Notice that
\begin{equation}
\int dE\ P(E;m_1,\cdots,m_n)=\left({L\over a}\right)^{2n}.
\end{equation}
This scaling can be calculated in closed form for $n=1$:
\begin{eqnarray}
&&
P(E;m)=
{
\exp\left[\ln\left({L\over a}\right)^2-
{\left(E-\overline{H_{1}}\right)^2\over4
Jg_A\overline{H_1}}\right]
\over\sqrt{4\pi Jg_A\overline{H_1} }
},
\\
&&
\overline{H_1}=m^2\pi J\ln\left({L\over a}\right).
\end{eqnarray}
Here, the arbitrary length scale $l_0$ is chosen to be $L$.

In the absence of disorder, the ground state of Eq. (\ref{H{1...n}})
is the ferromagnetic state $\tilde\Theta=0$.
In the presence of disorder, we can calculate the probability
$P(E;{\bf x}_1,\cdots,{\bf x}_n)$ from Eq. (\ref{formula for P(E; x1...xn)})
to find a vortex configuration with energy $E<0$.
We say that it is energetically favorable to create a vortex
configuration $\tilde\Theta$ different from $\tilde\Theta=0$
if $P(E<0;{\bf x}_1,\cdots,{\bf x}_n)$
does not vanish.
For example, to a good approximation, we find that
\begin{eqnarray}
\int_{-\infty}^0dE\ P(E;m)&&\approx\ 
P(0;m)
\nonumber\\
&&=
{
\left({L\over a}\right)^{2\left(1-{\pi m^2\over8g_A}\right)}
\over\sqrt{4\pi^2 J^2g_Am^2\ln(L/a) }
}.
\label{estimate P(0,m)}
\end{eqnarray}
Equation (\ref{estimate P(0,m)}) gives the number of sites
on which it is energetically favorable to create a single vortex
configuration.

As proposed in \cite{Fertig-Nattermann-Scheidl-Tang}, 
one can use estimate Eq. (\ref{estimate P(0,m)})
to establish a {\it  criterion} for the destruction
of long-range order at $T=0$ of the pure system 
by the random vector potential
$\tilde\partial_\mu\theta$. 
Long-range order is destroyed if the number of sites on which
it is energetically favorable to create a single vortex configuration
diverges in the thermodynamic limit 
$L\uparrow\infty$, i.e., if
the variance $g_A$ is larger than the critical value
\begin{equation}
(g_A)_{\rm crit}:={\pi\over8}m^2=\left({m\over2}\right)^2\times{\pi\over2}.
\end{equation}

It is important to stress that this condition does not guaranty
the existence of quasi-long-range order at $T=0$ if
$g_A<(g_A)_{\rm crit}$. To see this one can estimate
the number of pairs of sites on which it is energetically
favorable to create a dipole. This number is approximately
$P(0,+m,-m)$. In turn, with the choice $l_0=a$, $P(0,{\bf x}_1,{\bf x}_2)$ 
can be read off from $P(0,{\bf x}_1)$ provided one replaces 
$\ln(L/a)$ in $P(0,{\bf x}_1)$ by $2\ln(|{\bf x}_1-{\bf x}_2|/a)$:
\begin{eqnarray}
&&
P(0,+m,-m)=
\nonumber\\
&&
{4\pi\over\sqrt{8\pi^2J^2g_Am^2}}\left({L\over a}\right)^2
\int_0^{\sqrt{\ln(L/a)}}dy\ e^{2\left(1-{\pi m^2\over4g_A}\right)y^2}.
\end{eqnarray}
Hence, the number of pairs of sites on which it is energetically
favorable to create a dipole {\it always} diverges with system size 
[like $(L/a)^2$ if $g_A<2(g_A)_{\rm crit}$, like
$(L/a)^{4-(\pi m^2)/(2 g_A)}$ otherwise].
Likewise, the number of $n$ sites on which it is energetically favorable
to create a {\it neutral} vortex configuration always diverge with system size.
Hence, for most realizations of the disorder $\tilde\partial_\mu\theta$,
the ground states are not the ferromagnetic state $\tilde\Theta=0$
but are neutral and non-trivial ({\it complex}) vortex configurations.
To decide whether such complex ground states support quasi-long-range
order, one must establish absence of screening of the CB potential
between to external charges. 

\subsection{Screening at zero temperature}
\label{Screening at zero temperature}

To illustrate the issue of screening,
we introduce the random energy
\begin{equation}
H_{\rm CB}[\tilde\Theta,\theta_{\alpha}]:= 
-\pi J\sum_{k,l}(m_k-n_k) G_{kl} (m_l-n_l) ,
\label{toy model 1}
\end{equation}
where $\tilde\Theta$ is a neutral configuration of charges
$m_k=0,\pm1$,
$\theta_{\alpha}$ is a neutral configuration of charges
$n_k=0,\pm\alpha$, $0<\alpha<1$, that realizes the disorder, 
and $G_{kl}$ is a short hand notation
for the logarithmic CB gas potential.
Finally, we take the probability to realize $\theta_{\alpha}$ to be
proportional to
\begin{equation}
\exp\left[-{\pi\over g_A} \sum_{k,l}n_k G_{kl}n_l\right].
\label{toy model 2}
\end{equation}
The relationship between this model and the random phase $XY$ model
is that only a very small subset of all possible realizations
of the disorder in the random phase $XY$ model are allowed in 
Eqs. (\ref{toy model 1},\ref{toy model 2}). Thrown out are all
realizations of the disorder made of vortices of unequal
vorticity.

Consider now the vortex configuration $\Xi_{\alpha}$ defined by 
$m_k=0,\pm1$, respectively, 
whenever $n_k=0,\pm\alpha$. The energy of this configuration,
whose unit vortices track the fractionally charged quenched vortices, is
\begin{equation}
H_{\rm CB}[\Xi_{\alpha},\theta_{\alpha}]:=
-\pi J(1-\alpha)^2\sum_{k,l} m_k G_{kl} m_l, 
\end{equation}
and should be compared to the energy of the ferromagnetic state
\begin{equation}
H_{\rm CB}[0,\theta_{\alpha}]:=
-\pi J\alpha^2\sum_{k,l} m_k G_{kl} m_l. 
\end{equation}
We see that it is energetically favorable to create a unit vortex $m_k$
at the location of every quenched vortex $n_k$ provided 
$1/2<\alpha<1$. Otherwise, the ferromagnetic state is preferred.
For $\alpha=1/2$, $\Xi_{\alpha}$ and $\theta_{\alpha}$ are degenerate
\footnote{
We are indebted to E. Fradkin for this observation.}.

We are in position to ask the following question.
Is the CB potential screened or not for the family of vortex
configurations $\{\Xi_{\alpha}\}$ where $1/2<\alpha<1$ is held fixed?
Since $\Xi_{\alpha}$ merely creates vortices wherever quenched vortices sit, 
the question can be reduced to: what are the screening properties
of the CB gas with effective temperature and charge
$g_A$ and $\alpha$, respectively? 
The answer is known \cite{KT transition}, 
namely for sufficiently small $g_A$, i.e.,
\begin{equation}
g_A<\alpha^2{\pi\over2},
\end{equation}
the CB gas does not screen since the dipole phase is realized and
there exists quasi-long-range order.
Otherwise, the CB gas screens since the plasma phase is realized
and quasi-long-range order is destroyed.

Once we know the asymptotic form of the CB potential between two
external charges for the family $\{\Xi_{\alpha}\}$, 
$1/2<\alpha<1$ and  $g_A$ fixed, 
we can extract the $\alpha$-averaged CB potential 
where we restrict $1/2<\alpha<1$.
The  $\alpha$-averaged CB potential is controlled in the thermodynamic limit
by $\alpha=1/2$ since the probability in Eq. (\ref{toy model 2}) scales like
$\exp[-\alpha^2 (L/a)^2f(L/a)]$, where $f(x)$ is some positive function
with $\lim_{x\uparrow\infty} f(x)/x^2=0$ that does not depend on $\alpha$.
The $\alpha$-averaged CB potential, $1/2<\alpha<1$,
thus decays logarithmically for sufficiently small $g_A$
and decays exponentially
if $g_A>(g_A)_{\rm crit}$. Hence, quasi-long-range order is present for not
too strong disorder strength provided it can be shown that
$\Xi_{\alpha}$ is the ground state for every realizations of the disorder
when $1/2<\alpha<1$. 

Although this conjecture might be reasonable for
the toy model, it is certainly not true for the full problem
where a given realization of the disorder $\tilde\partial_\mu\theta$ 
can create vortices 
of arbitrary fractional charges in contrast to the toy model.
Hence, although the existence of both $\theta_\alpha$ and $\Xi_{\alpha}$
suggests the existence of $(g_A)_{\rm crit}$ in the full model,
we cannot rule out more complex configurations $\tilde\Theta$
with energies lower than that of both the ferromagnetic state and $\Xi_\alpha$
and for which the CB potentials are screened at long distances for some 
$g_A<(g_A)_{\rm crit}$.

\section{Fugacity expansion}
\label{Fugacity expansion}

In this section, 
we we are going to expand all correlation functions of the SG operator
$e^{{\rm i}\chi}$ in powers of $h_1/2t$
and then perform disorder averaging order by order
in powers of $h_1/2t$. We thus reproduce the fugacity
expansion performed by Korshunov \cite{Korshunov 1993} on the
CB gas with quenched fractionally charged vortices. 
Scaling fields with negative scaling dimensions are then seen to control
the singular behavior of the expansion in powers of $h_1/2t$
of the inverse SG partition function order by order.
However, we begin first by illustrating a possible drawback of the
fugacity expansion.

\subsection{Drawback of the fugacity expansion}
\label{Drawbacks of the fugacity expansion}

The expansion of the inverse SG partition function in powers of
the fugacity $h_1/2t$ should be taken with great caution since convergence
is not always warranted. Indeed,
let $X$ be a real valued random variable and let $Y:=1/(1+X)$
be another real valued random variable. For example, $Y$ could be
$1/{\cal Z}_{\rm CB}$ whereby $X$ could be ${\cal Z}_{\rm CB}-1$.
Let $P_X(x)$ be the probability that $X$ takes the value $x$.
We want to calculate the probability $P_Y(y)$ that $Y$ takes
the value $y$.  For definiteness,
\begin{itemize}

\item Case I:
\begin{eqnarray}
&&
P_X(x):=\ e^{-x},\quad 0\leq x<\infty\Leftrightarrow
\nonumber\\
&&
P_Y(y)=\ {e^{1-{1\over y}}\over y^2},\quad 0\leq y\leq 1.
\label{P_X}
\end{eqnarray}

\item Case II:
\begin{eqnarray}
&&
P_X(x):=\ x^{-2},\quad 1\leq x<\infty\Leftrightarrow
\nonumber\\
&&
P_Y(y)=\ (1-y)^{-2},\quad 0\leq y\leq {1\over2}.
\label{P_Y}
\end{eqnarray}

\end{itemize}
We immediately conclude that the expansion 
\begin{equation}
Y=\sum_{n=0}^\infty{(-1)^n\over n}X^n,
\end{equation}
would predict that all moments of $Y$ diverge in both cases I and II. 
However, these moments can be calculated directly 
from Eqs. (\ref{P_X},\ref{P_Y}) and are all finite.

\subsection{Preview to the fugacity expansion to fourth order}
\label{subsec:Preview to the fugacity expansion to fourth order}

Let $n$ be a positive integer,
choose $n$ points ${\bf x}_1$, $\cdots$, ${\bf x}_{n}$ 
on the Euclidean plane, and define
\begin{eqnarray}
&&
F_{{\bf x}_1,\cdots,{\bf x}_{n}}:= 
e^{{\rm i}\varepsilon_1\chi({\bf x}_1)}\cdots 
e^{{\rm i}\varepsilon_{n}\chi({\bf x}_{n})},
\label{F(1 to n)}
\\
&&
\langle F_{{\bf x}_1,\cdots,{\bf x}_{n}}\rangle^{\rm unnor}:=
\int\! {\cal D}[\chi]\ e^{-S_{\rm SG}[\chi,\theta]}
F_{{\bf x}_1,\cdots,{\bf x}_{n}}.
\label{F(1 to n) disc}
\end{eqnarray}
Each factor $e^{{\rm i}\varepsilon_k\chi({\bf x}_k)}$, $k=1,\cdots,n$, 
can be thought of as the insertion of an external vortex of vorticity
$\varepsilon_k=\pm 1$ in the CB gas. 
Thermal averaging of Eq. (\ref{F(1 to n)}) is obtained by 
dividing the unnormalized average in Eq. (\ref{F(1 to n) disc})
by the SG partition function:
\begin{equation}
\langle F_{{\bf x}_1,\cdots,{\bf x}_{n}}\rangle=
{\langle F_{{\bf x}_1,\cdots,{\bf x}_{n}}\rangle^{\rm unnor}\over 
{\cal Z}_{\rm SG}[\theta]}.
\label{F(1 to n) conn}
\end{equation}
Finally, disorder averaging over $\tilde\partial_\mu\theta$ 
is done with the probability 
distribution of Eq. (\ref{P[theta,eta]}). 

We attempt to
calculate both $\langle F_{{\bf x}_1,\cdots,{\bf x}_{n}}\rangle$ 
and $\langle F_{{\bf x}_1,\cdots,{\bf x}_{n}}\rangle^{\rm unnor}$ 
through a power expansion in $h_1/2t$ and then perform term by term
disorder averaging over $\tilde\partial_\mu\theta$. 
Thermal averaging and $\tilde\partial_\mu\theta$ 
averaging are seen to ``factorize'' 
to each order in $h_1/2t$ for
$\langle F_{{\bf x}_1,\cdots,{\bf x}_{n}}\rangle^{\rm unnor}$.
We then go on performing $\tilde\partial_\mu\theta$ averaging over 
$\langle F_{{\bf x}_1,\cdots,{\bf x}_{n}}\rangle$.

The key identity that is needed is 
\begin{eqnarray}
&&
\langle F_{{\bf x}_1,\cdots,{\bf x}_{n}}\rangle^{\rm unnor}=
\\
&&
\sum_{m=0}^{\infty}
{1\over m!}\left({h_1\over2t}\right)^m
\sum_{p=0}^m\pmatrix{m\cr p\cr}
\int d^2{\bf y}_1\cdots\int d^2{\bf y}_m\times
\nonumber\\
&&
\left\langle 
e^{{\rm i}\chi({\bf y}_1)}\cdots
e^{{\rm i}\chi({\bf y}_p)}e^{-{\rm i}\chi({\bf y}_{p+1})}\cdots
e^{-{\rm i}\chi({\bf y}_m)}
F_{{\bf x}_1,\cdots,{\bf x}_{n}}
\right\rangle^{\rm unnor}_{h_1=0}.\nonumber
\end{eqnarray}
Thermal averaging on the last line must be performed with $h_1=0$,
in which case averaging over $\chi$ is Gaussian. We can then use the 
shift of integration variable $\chi=:\chi'+{{\rm i}t\over2\pi}\theta$
to decouple averaging over $\chi'$ from averaging over 
$\tilde\partial_\mu\theta$:
\begin{eqnarray}
&&
\langle F_{{\bf x}_1,\cdots,{\bf x}_{n}}\rangle^{\rm unnor}=
\label{factorization}
\\
&&
\sum_{m=0}^{\infty}
{1\over m!}\left({h_1\over2t}\right)^m
\sum_{p=0}^m\pmatrix{m\cr p\cr}
\int d^2{\bf y}_1\cdots\int d^2{\bf y}_m\times
\nonumber\\
&&
e^{-{t\over2\pi}\theta({\bf y}_1)}\cdots
e^{{t\over2\pi}\theta({\bf y}_m)}
e^{-{t\over2\pi}\sum_{k=1}^{n}
\varepsilon_k\theta({\bf x}_k)}\times
\nonumber\\
&&
\left\langle 
e^{{\rm i}\chi'({\bf y}_1)}\cdots
e^{-{\rm i}\chi'({\bf y}_m)}
e^{{\rm i}\sum_{k=1}^{n}\varepsilon_k\chi'({\bf x}_k)}
\right\rangle^{\rm unnor}_{h_1=0}.\nonumber
\end{eqnarray}

Thus to each order in $h_1/2t$,
averaging over $\tilde\partial_\mu\theta$ 
has factorized from averaging over $\chi'$
in the integrand on the right hand side of Eq. (\ref{factorization}).
In fact since both averages are Gaussian, 
one obtains the CB gas representation 
\begin{eqnarray}
&&
\overline{ \langle F_{{\bf x}_1,\cdots,{\bf x}_{n}}\rangle^{\rm unnor} }=
\nonumber\\
&&
\sum_{m=0}^{\infty}
{1\over m!}\left({h_1\over2t}\right)^m
\sum_{p=0}^m\pmatrix{m\cr p\cr}
\int d^2{\bf y}_1\cdots\int d^2{\bf y}_m\times
\nonumber\\
&&
\exp\left[+\pi \bar K\sum_{k,l=1}^{m+n}\varepsilon_k\varepsilon_l
\ln\left|{{\bf z}_k-{\bf z}_l\over l_0}\right|\right],
\label{overline F(1 to n)}
\end{eqnarray}
where
\begin{equation}
{\bf z}_k=
\cases{
{\bf y}_k,&if $k=1  ,\cdots, m$,\cr
{\bf x}_k,&if $k=m+1,\cdots, m+n$,\cr
}
\end{equation}
and we have introduced the effective CB gas coupling constant
\begin{equation}
2\pi\bar K:= 
2\pi(K- g_AK^2)=
2\pi\left[ 
{t\over 4\pi^2}+{g_A\over 4\pi^2}\times\left({\rm i}{t\over2\pi}\right)^2
\right].
\label{bar K}
\end{equation}

Equations (\ref{overline F(1 to n)},\ref{bar K}) tell us that
under the assumption that
a fugacity expansion of the CB gas is valid for
each realizations of the disorder $\tilde\partial_\mu\theta$,
then averaging over $\tilde\partial_\mu\theta$ 
all unnormalized correlation functions
consisting in the insertion of external charges
amounts to the {\it same} renormalization of the
CB gas effective temperature $1/\pi K$. Moreover, in view of
\begin{eqnarray}
&&
{\cal Z}_{\rm SG}[\theta]=
\nonumber\\
&&
\sum_{m=0}^\infty{1\over m!}\left({h_1\over2t}\right)^m
\sum_{p=0}^m\pmatrix{m\cr p\cr}\int d^2{\bf y}_1\cdots d^2{\bf y}_m\times
\nonumber\\
&&
\left\langle 
e^{{\rm i}\chi({\bf y}_1)}\cdots 
e^{{\rm i}\chi({\bf y}_{p})}e^{-{\rm i}\chi({\bf y}_{p+1})}\cdots
e^{-{\rm i}\chi({\bf y}_m)}
\right\rangle^{\rm unnor}_{h_1=0},
\end{eqnarray}
this is also true for the case $n=0$ which corresponds to the
average of the partition function expanded in powers of the fugacity.
We thus expect that the fugacity expansion for all
unnormalized correlation functions 
$\overline{\langle F_{{\bf x}_1,\cdots,{\bf x}_{n}}\rangle^{\rm unnor}}$,
if well defined, is convergent in the same region of the $1/K$, $g_A$, 
$h_1/2t$
coupling space. However, this is not so for the normalized average
$\overline{\langle F_{{\bf x}_1,\cdots,{\bf x}_{n}}\rangle}$.

Indeed Eqs. (\ref{overline F(1 to n)},\ref{bar K}) are not sufficient
to extract the effect of disorder averaging over $\tilde\partial_\mu\theta$ 
on the perturbative expansion in powers of $h_1/2t$ 
of $\langle F_{{\bf x}_1,\cdots,{\bf x}_{n}}\rangle$.
In fact, we need {\it all moments} (not only the first one) of the unnormalized
correlation functions
$\langle F_{{\bf x}_1,\cdots,{\bf x}_{n}}\rangle^{\rm unnor}$
as can be seen by expanding $1/{\cal Z}_{\rm SG}[\theta]$ in powers of 
$h_1/2t$ in Eq. (\ref{F(1 to n) conn}). 
But these moments can be estimated from
\begin{eqnarray}
&&
\overline{ 
\left[
\left\langle 
e^{{\rm i}\chi({\bf y}_1)}\cdots 
e^{{\rm i}\chi({\bf y}_{p})}e^{-{\rm i}\chi({\bf y}_{p+1})}\cdots
e^{-{\rm i}\chi({\bf y}_m)}
\right\rangle^{\rm unnor}_{h_1=0}
\right]^q }=
\nonumber\\
&&
\exp\left[+\pi \overline{K(q)}\sum_{k,l=1}^{m}\varepsilon_k\varepsilon_l
\ln\left|{{\bf y}_k-{\bf y}_l\over l_0}\right| 
\right],
\label{these moments}
\end{eqnarray}
where
\begin{equation}
\overline{K(q)}:= 
Kq- g_A(Kq)^2.
\label{bar K(q)}
\end{equation}
The boundary $\overline{K(q)}=0$ which delimits a positive from a negative
effective CB gas temperature shrinks with increasing $q$. Hence
order by order in powers of $h_1/2t$, the regime in which the fugacity
expansion is well defined is controlled by the largest moment 
contributing to this order in Eq. (\ref{these moments}). 
We recognize in Eq. (\ref{bar K(q)}) the scaling dimensions on the right hand
side of Eq. (\ref{most relevant}). 


\subsection{CB gas interpretation of the fugacity expansion}
\label{CB gas interpretation of the fugacity expansion}

Before going into a more detailed discussion of the fugacity expansion up to
fourth order, we comment on some general features of the fugacity expansion
and its close relationship to the replicated effective theory of
section \ref{The perturbative instability}.
The impurity average over the two-point correlation function 
$\langle F_{12}\rangle$ in Eq. (\ref{moments 2 point})
can be recast as a summation over all possible configurations
\bleq
\ifpreprintsty\else \renewcommand{\thesection}{\Alph{section}} %
\renewcommand{\theequation}{\Alph{section}\arabic{equation}} \fi %
\begin{equation}
\overline
{ 
\langle 
e^{{\rm i}\chi({\bf x}_1)       -{\rm i}\chi( {\bf x}_2  )}
\times
e^{{\rm i}\chi({\bf y}_1)+\cdots-{\rm i}\chi({\bf y}_{2i})} 
\rangle^{\rm unnor}_{h_1=0}
\times
\langle  
e^{{\rm i}\chi({\bf z}_1)+\cdots-{\rm i}\chi({\bf z}_{2j})}
\rangle^{\rm unnor}_{h_1=0}
\times
\cdots
}
\end{equation}
of $2i+2j+\cdots$ vortices that screen the two external vortices at
${\bf x}_1$ and ${\bf x}_2$.  
The appropriate combinatorial weight that results from 
expanding the numerator and denominator in powers of the fugacity 
is not written here but can be found in the coming subsection.
The first thermal average $\langle\cdots\rangle^{\rm unnor}_{h_1=0}$ 
comes from the
expansion of the numerator in Eq. (\ref{RG of correlation function}), 
while the remaining factors in 
$\langle\cdots\rangle^{\rm unnor}_{h_1=0}\times\cdots$ 
come from the expansion of the denominator. 
Notice that thermal averages are disconnected averages
as is indicated by the superscript
[see Eq. (\ref{F(1 to n) disc})]. 

Introducing $\chi=:\chi'+\frac{it}{2\pi}\theta$ (since $\chi'$ and $\theta$
decouple), the above equation becomes
\begin{equation}
\overline
{ 
e^{-{t\over2\pi}\theta({\bf x_1})+\cdots+{t\over2\pi}\theta({\bf z}_{2j})+
\cdots}
\langle 
e^{{\rm i}\chi'({\bf x}_1)       -{\rm i}\chi'( {\bf x}_2  )}
\times
e^{{\rm i}\chi'({\bf y}_1)+\cdots-{\rm i}\chi'({\bf y}_{2i})} 
\rangle^{\rm unnor}_{h_1=0}
\times
\langle  
e^{{\rm i}\chi'({\bf z}_1)+\cdots-{\rm i}\chi'({\bf z}_{2j})}
\rangle^{\rm unnor}_{h_1=0}
\times
\cdots
}.
\label{decoupled generic term in fugacity expansion}
\end{equation}
\eleq
\noindent
Each term in the fugacity expansion can then be given the 
following CB gas interpretation.
There is one charge $Q$, {\it the disorder charge}, 
associated with the field $\theta$. It is also
convenient to associate with each disconnected 
thermal average $\langle\cdots\rangle^{\rm unnor}_{h_1=0}$ appearing in 
(\ref{decoupled generic term in fugacity expansion})
a distinct thermal charge $\varepsilon_a$.
More precisely, to order $(h_1/2t)^{2n}$ we introduce
$n+1$ thermal charges $\varepsilon_a$, $a=0,\cdots,n$
(we take the thermal charges $\varepsilon_{m+1}=\cdots=\varepsilon_n=0$
to always vanish if there are only $m<n$ thermal factors,
and the charges labeled by the subscript 0 always refer to the disconnected
thermal average involving the two external charges at ${\bf x}_1$
and ${\bf x}_2$, respectively). 
For example, if none of the coordinates in
(\ref{decoupled generic term in fugacity expansion}) coincide we assign
\begin{itemize}
\item
$(\varepsilon_0=+1,\varepsilon_1=0,\cdots,\varepsilon_n=0;Q=+1)$ 
to the CB charges at ${\bf x}_1$ and ${\bf y}_l$, $l=1,  \cdots,i $.
\item
$(\varepsilon_0=-1,\varepsilon_1=0,\cdots,\varepsilon_n=0;Q=-1)$ 
to the CB charges at ${\bf x}_2$ and ${\bf y}_l$, $l=i+1,\cdots,2i$.
\item
$(\varepsilon_0=0,\varepsilon_1=+1,\cdots,\varepsilon_n=0;Q=+1)$ 
to the CB charges at ${\bf z}_l$, $l=1  ,\cdots, j$.
\item
$(\varepsilon_0=0,\varepsilon_1=-1,\cdots,\varepsilon_n=0;Q=-1)$ 
to the CB charges at ${\bf z}_l$, $l=j+1,\cdots,2j$.
\end{itemize}

Since coordinates belonging to
{\it distinct} disconnected thermal averages can coincide
(see the next subsection),
we will also allow {\it complex or fused} charges of the form 
$(\varepsilon_0,\cdots,\varepsilon_n;Q)$
where
\begin{equation}
\sum_{a=0}^n \varepsilon_n=Q,\quad 
|\varepsilon_1|\leq n,\quad 
|\varepsilon_2|\leq n-|\varepsilon_1|,\cdots.
\end{equation}
It is then possible to systematically carry through the thermal and disorder
average in (\ref{decoupled generic term in fugacity expansion})
by assuming that the {\it replicated} charges
$(\varepsilon_0,       \varepsilon_1,       \cdots;Q       )$ and 
$(\varepsilon_0^\prime,\varepsilon_1^\prime,\cdots;Q^\prime)$
are associated to operators with correlation functions given by
\begin{equation}
\left|{{\bf w} -{\bf w'}\over a}\right|^
{-2\pi g_A K^2 Q Q^\prime+2\pi K \sum_a
\varepsilon^{\ }_a \varepsilon^\prime_a}.
\end{equation}

As we change the microscopic cut-off $a$ and/or include randomness in $h_1$, 
two operators with charges 
$(\varepsilon_0,       \varepsilon_1,       \cdots;Q       )$ and 
$(\varepsilon_0^\prime,\varepsilon_1^\prime,\cdots;Q^\prime)$,
respectively, may fuse into one with charge 
$
(\varepsilon_0^{\ }+\varepsilon_0^\prime,
\varepsilon_1^{\ }+\varepsilon_1^\prime,\cdots;Q+Q^\prime)
$. 
Thus, operators with arbitrary complex charges may appear in the 
fugacity expansion upon renormalization or averaging over random
vortex fugacity. 
Although bare vortices in the $XY$ model have a simple structure (being
labeled by a single integer), it is striking to see that 
complex vortices {\it must} be accounted for in disorder averaged 
correlation functions in contrast to the clean limit
(in the pure system these higher charges are always irrelevant). 
Complex vortices also appear formally in the replica approach
(see section \ref{The perturbative instability}). However, 
intuition is easily lost when taking the replica limit $r\downarrow 0$.

Our discussion of the fugacity expansion suggests that fugacities
$Y_{(\varepsilon_0,\varepsilon_1,\cdots;Q)}$ should be introduced when
operators with charge $(\varepsilon_0,\varepsilon_1,\cdots;Q)$ contribute to
(\ref{decoupled generic term in fugacity expansion}).
These fugacities, to a first approximation, are related to the density
of {\it screening} charges of type $(\varepsilon_0,\varepsilon_1,\cdots;Q)$.
Form invariance of the fugacity expansion under an infinitesimal
rescaling of the short distance cutoff $a$:
\begin{equation}
a':= ae^l,\quad 0<l\ll1,
\end{equation}
would then imply that the fugacities obey RG equations.  
Since the scaling dimension of an operator depends on its charge, 
we expect the RG equations for the new fugacities to be different 
from each other. In particular, to order $2n$ in the fugacity, 
the generalized fugacities 
$Y_{(0,1,\cdots,1;Q)},\cdots,Y_{(1,\cdots,1,0;Q)}$ are expected to be
the most relevant operators within the family 
$Y_{(\varepsilon_0,\cdots,\varepsilon_n;Q)}$
[compare with Eq. (\ref{most relevant})].

We will slightly abuse our notation in the following by using
for the subscript of generalized fugacities
the maximum number $n$ of non-vanishing thermal charges 
(to order $2n$ in the fugacity expansion). 
For example, to fourth order, we will denote by $Y_{(1,1;2)}$ 
any of the three fugacities
$Y_{(1,1,0;2)}$, $Y_{(1,0,1;2)}$, and $Y_{(0,1,1;2)}$
associated to the charges 
$(1,1,0;2)$, $(1,0,1;2)$, and $(0,1,1;2)$, respectively.
Indeed, we will show that the RG equations for all 
three fugacities are the same.

We illustrate those general considerations with a detailed calculation
of the fourth order correction to 
$\overline{\langle F_{{\bf x}_1,{\bf x}_2}\rangle}$.
We denote with $\overline{F^{(4)}_{12}}$
the fourth order coefficient to the fugacity expansion
(see the following subsection).
It is then convenient to distinguish between three contributions to
$\overline{F^{(4)}_{12}}$, denoted
$\overline{A^{\ }_{12}}$, 
$\overline{B^{\ }_{12}}$, and 
$\overline{C^{\ }_{12}}$, respectively.
To fourth order in the fugacity expansion, we expect 
that complex vortices with charges 
$(\varepsilon_0,\varepsilon_1,\varepsilon_2;Q)$ emerge. 
Correspondingly, it should be possible to recast the RG equations in terms of 
fugacities 
$Y_{(\varepsilon_0,\varepsilon_1,\varepsilon_2;Q)}$. 
This is indeed so.

The contribution $\overline{A^{\ }_{12}}$ is nothing but the
second order contribution squared. Hence, it is due to the screening
of two external vortices 
with charges $(+1;+1)$ and $(-1;-1)$, respectively,
by four thermal vortices with charges of type $(\pm1;\pm1)$.
It is given by:
\begin{eqnarray}
\overline{A^{\ }_{12}}\left({h_1\over2t}\right)^4\approx
\left|{{\bf x}_{12}\over a}\right|^{-2\pi\bar K}\times
{1\over2}\left[2\pi x^{(2)}\ln\left|{{\bf x}_{12}\over a}\right|\right]^2,
\label{A_{12}}
\end{eqnarray}
where $x^{(2)}$ is the second order correction of 
Eq. (\ref{x(q)}) when $q=1$:
\begin{eqnarray}
&&x^{(2)}:=
- 4\pi^3\left[ \bar K^2  
- K^4 g^2_A  \right]Y^2_{(1;1)}\!
\int\limits_1^{L/a}\!\!\! d y y^{3-2\pi\bar K}.
\end{eqnarray}
This is an encouraging result since it justifies a posteriori 
our assumption in Eq. (\ref{algebraic moment for F_12 to 2 order})
that logarithmic corrections can be reexponentiated.

The contribution $\overline{B^{\ }_{12}}$ is due to the screening
of our two external vortices 
by either three replicated vortices with charges of type
$(+1,+1;+2)$ (fused), 
$(-1,0 ;-1)$ and 
$(0,-1 ;-1)$, respectively,
or three replicated vortices with charges
$(+1,-1;0 )$ (fused), 
$(-1,0 ;-1)$ 
and 
$(0,+1 ;+1)$, respectively.

The term $\overline{C^{\ }_{12}}$ is induced by the screening of  
two external vortices at ${\bf x}_1$ and ${\bf x}_2$, respectively,
by either two fused vortices with charges of type
$(\pm1,\mp1;0)$,
or by two fused vortices with charges of type
$(\pm1,\pm1;\pm2)$.
Concretely, we find
\begin{eqnarray}
\overline{C^{\ }_{12}}\left({h_1\over2t}\right)^4\approx
\left|{{\bf x}_{12}\over a}\right|^{-2\pi\bar K}\times
2\pi x^{(4)}_c\ln\left|{{\bf x}_{12}\over a}\right|,
\label{C_{12}}
\end{eqnarray}
where $x^{(4)}_c$ is the fourth order correction due to two-body
renormalization effects:
\begin{eqnarray}
x^{(4)}_c&&:=
-4\pi^3 K^2
Y^2_{(1,-1;0)}\!
\int\limits_1^{L/a}\!\! dy y^{3-4\pi K}
\label{x(4)_c}
\\
&&+
4\pi^3\left\{4K^4g^2_A-{\left[\overline{K(2)}\right]^2\over4}\right\}
Y^2_{(1,1;2)}\!\int\limits_1^{L/a}\!\! dyy^{3-2\pi\overline{K(2)}}.
\nonumber
\end{eqnarray}

The first line on the right hand side
of Eq. (\ref{x(4)_c}) is fully oblivious of the disorder.
The second line is dramatically sensitive to it since it yields a new boundary
for analyticity of the fugacity expansion. 
The two lines are equal in the absence of disorder.

Furthermore, the second line on the right hand side of 
Eq. (\ref{x(4)_c}) implies that,
in contrast to the second order contribution to the fugacity expansion
of the mean $\overline{\langle F_{12}\rangle}$, the scaling exponent
$2\pi\overline{K(2)}$ of $\overline{\langle F_{12}\rangle^2_{h_1=0}}$
enters in the fourth order correction to the scaling
exponent $2\pi\bar K$ of $\overline{\langle F_{12}\rangle^{\ }_{h_1=0}}$. 
Hence, the second moment of  $\langle F_{12}\rangle$ couples
to the mean of $\langle F_{12}\rangle$ 
in the RG flow to fourth order in the fugacity.

{ }From the scaling dimension $2\pi K- 4\pi g_A K^2=\pi \overline{ K(2)} $ 
of the operator with charge of type $(1,1;2)$,
we find the RG equation (\ref{dY112}) of $Y_{(1,1;2)}$. 
Equation (\ref{x(4)_c}) tells us that $Y_{(1,1;2)}$ 
indeed enters the RG equation of $\bar K$. 
Hence, the perturbative RG equations breaks down
when $2\pi \overline{ K(2)}  < 4$.

The pattern for renormalization should become clear 
from our fourth order calculation. To each new order $2n$ in $h_1/2t$
we need to introduce new fugacities for replicated vortices
labeled by their thermal and disorder charges 
\begin{eqnarray}
(\varepsilon_0,\cdots,\varepsilon_n;\sum^n_{a=1}\varepsilon_a),
\quad \varepsilon_a=\pm1,\quad a=0,\cdots,n,
\nonumber
\end{eqnarray}
to  close the RG equations. The scaling dimension of the fugacity
\begin{eqnarray}
Y^{\ }_{(\varepsilon_1,\cdots,\varepsilon_n;\sum^n_{a=1}\varepsilon_a)}
\nonumber
\end{eqnarray}
on the critical plane of Fig. 1 is deduced from that of 
\begin{eqnarray}
&&
\exp\left[{\rm i}\sum_{a=1}^n\varepsilon^{\ }_a\chi^{\ }_a({\bf y})\right]=
\nonumber\\
&&
\exp\left[{\rm i}\sum_{a=1}^n\varepsilon^{\ }_a\chi'    _a({\bf y})\right]
\exp\left[-2\pi K\theta({\bf y})\sum_{a=1}^n\varepsilon^{\ }_a\right].
\end{eqnarray}

For any infinitesimal value of the disorder strength $g_A$,
the contribution to the coefficient of the fugacity expansion to order 
$2n$ that defines the regime of analyticity [see Eq. (\ref{g^(n)})] describes 
the screening of the CB interaction by two tightly bound fused vortices
of charges $(\varepsilon,\cdots,\varepsilon;n\varepsilon)$, $\varepsilon=\pm1$,
respectively. The scaling with system size of this coefficient is thus given by
Eq. (\ref{most divergent integral to order 2n}). 
Subleading contributions
to the coefficient of the expansion originate from screening of the CB 
interaction by three and more replicated vortices. 

Finally, we would like
to stress that there exists a one to one correspondence between the
replicated vortices appearing in the fugacity expansion and the 
replicated primary fields constructed in section 
\ref{The perturbative instability} 
[see Eqs. 
(\ref{cumulant expansion if random core},\ref{most and less relevant})]
as was first suggested by Korshunov \cite{Korshunov 1993}.

\subsection{Two-point function
$\overline{\langle F_{{\bf x}_1,{\bf x}_2}\rangle}$
up to order $(h_1/2t)^4$}
\label{Two-point function to fourth order in h_1/2t}

\bleq
\ifpreprintsty\else \renewcommand{\thesection}{\Alph{section}} %
\renewcommand{\theequation}{\Alph{section}\arabic{equation}} \fi %

It is very instructive to carry the fugacity expansion of 
the two-point function
\begin{eqnarray}
&&
F_{{\bf x}_1,{\bf x}_2}:= 
e^{{\rm i}\chi({\bf x}_1)-{\rm i}\chi({\bf x}_2)}\equiv F_{12}.
\label{F(1 to 2)}
\end{eqnarray}
This calculation is the crucial ingredient in the RG analysis of the KT
transition. 

For the pure case quasi-long-range order holds
if $F_{12}$ decays algebraically with separation with a scaling exponent
$\kappa\equiv x(1)$ [see Eq. (\ref{algebraic moment for F_12 to 2 order})]
which is an analytic function of fugacity $Y_1$ in the vicinity of
$Y_1=0$. 
The transition temperature to a disordered phase is deduced from 
the boundary along the line $1/K\geq0$ for which the scaling exponent 
$\kappa$ becomes non-analytic.

In the presence of disorder, Rubinstein et al. performed 
the same analysis after averaging
over disorder the fugacity expansion of $F_{12}$ term by term
up to second order in $Y_1$. They inferred the parabolic boundary of Fig. 1
{ }from the onset of a non-analytic dependency of $\kappa$ on 
fugacity $Y_1$. We repeat their argument 
and show, in the spirit of Korshunov's analysis \cite{Korshunov 1993},
how higher moments of correlation functions 
invalidate their conclusion beyond second order in the fugacity expansion.

The power expansion in $Y_1\propto h_1/2t$ is given by
\begin{eqnarray}
\langle F_{12}\rangle&&:=
\sum_{n=0}^\infty F^{(n)}_{12}\ (h_1/2t)^n
:=
{\sum_{m=0}^\infty f^{(m)}_{12}\ (h_1/2t)^m\over
1+\sum_{n=1}^\infty Z^{(n)}\ (h_1/2t)^n},
\label{power expansion in fugacity of two-point function}
\end{eqnarray}
where $F^{(2n+1)}_{12}= f^{(2n+1)}_{12} = Z^{(2n+1)}=0$ and 
\begin{eqnarray}
&&
F^{(0)}_{12}=
f^{(0)}_{12},
\nonumber\\
&&
F^{(2)}_{12}= 
f^{(2)}_{12}-f^{(0)}_{12} Z^{(2)},
\\
&&
F^{(4)}_{12}= 
f^{(4)}_{12}-
\left[f^{(2)}_{12}Z^{(2)} + f^{(0)}_{12}Z^{(4)}\right]+
f^{(0)}_{12} Z^{(2)}\times Z^{(2)},
\nonumber
\end{eqnarray}
up to fourth order in $h_1/2t$.
The coefficients in the power expansions in $h_1/2t$ of the numerator and
denominator are
\begin{eqnarray}
&&
f^{(2n)}_{12}={1\over (n!)^2}
\int \underbrace{d^2{\bf y}_1\cdots d^2{\bf y}_{2n}}_{\neq}
\left\langle  
e^{ {\rm i}\chi({\bf y}_1    )+\cdots+{\rm i}\chi({\bf y}_n   )
   -{\rm i}\chi({\bf y}_{n+1})-\cdots-{\rm i}\chi({\bf y}_{2n})
   +{\rm i}\chi({\bf x}_1)-{\rm i}\chi({\bf x}_2)}
\right\rangle^{\rm unnor}_{h_1=0},
\\
&&
Z^{(2n)}={1\over (n!)^2}
\int \underbrace{d^2{\bf y}_1\cdots d^2{\bf y}_{2n}}_{\neq}
\left\langle  
e^{ {\rm i}\chi({\bf y}_1    )+\cdots+{\rm i}\chi({\bf y}_n   )
   -{\rm i}\chi({\bf y}_{n+1})-\cdots-{\rm i}\chi({\bf y}_{2n})}
\right\rangle^{\rm unnor}_{h_1=0},
\end{eqnarray}
respectively. 
Since we are assuming the existence of a dipole phase,
thermal vortices are taken with a hardcore as is indicated by the
constraint that the coordinates of the vortices cannot coincide. 
Hence, we are implicitly using
a short distance cutoff $a$ for the thermal vortices. 
As a matter of principle, this cutoff need
not be the same as that used at short distances for the quenched vortices.
Nevertheless, for notational simplicity, we will assume that quenched vortices
share the same hardcore radius.

To lowest order in $Y_1\propto h_1/2t$:
\begin{eqnarray}
\overline{F^{(0)}_{12}}=
\left|{{\bf x}_1-{\bf x}_2\over a}\right|^{-2\pi\bar K}\equiv
\left|{{\bf x}_{12}\over a}\right|^{-2\pi\bar K}.
\end{eqnarray}

To second order in $Y_1\propto h_1/2t$:
\begin{eqnarray}
\overline{F^{(2)}_{12}}&&=
\left|{a^2\over{\bf x}_1-{\bf x}_2}\right|^{2\pi\bar K}
\int \underbrace{d^2{\bf y}_1 d^2{\bf y}_2}_{\neq}
{
{\cal K}_{{\bf x}_1{\bf x}_2}({\bf y}_1,{\bf y}_2;2\pi\bar K)-
{\cal K}_{{\bf x}_1{\bf x}_2}({\bf y}_1,{\bf y}_2;2\pi K^2 g_A)
\over
|{\bf y}_1-{\bf y}_2|^{2\pi\bar K}
}
\nonumber\\
&&\equiv
\left|{a^2\over{\bf x}_{12}}\right|^{2\pi\bar K}
\underbrace{\int_{12}}_{\neq}
{
{\cal K}_{{\bf x}_1{\bf x}_2}({\bf y}_1,{\bf y}_2;2\pi\bar K)-
{\cal K}_{{\bf x}_1{\bf x}_2}({\bf y}_1,{\bf y}_2;2\pi K^2 g_A)
\over
|{\bf y}_{12}|^{2\pi\bar K}
}
\nonumber\\
&&\approx
|{\bf x}_{12}/a|^{-2\pi\bar K}\times 
8\pi^4\left(\bar K^2- K^4g^2_A\right)\times
\left(\int_a^{L}\! d|{\bf y}_{12}|\ 
|{\bf y}_{12}|^3|{\bf y}_{12}/a|^{-2\pi\bar K}\right)\times
\ln|{\bf x}_{12}/a|,
\label{F_{12}^{(2)}}
\end{eqnarray}
where
\begin{equation}
{\cal K}_{{\bf x}_1{\bf x}_2}({\bf y}_1,{\bf y}_2;x):=
\left[
{
|{\bf y}_1-{\bf x}_1||{\bf y}_2-{\bf x}_2|
\over
|{\bf y}_1-{\bf x}_2||{\bf y}_2-{\bf x}_1|
}
\right]^{x}.
\end{equation}
If we interpret the presence of a logarithmic correction on the right hand side
of Eq. (\ref{F_{12}^{(2)}}) as the first term in the expansion of
\begin{equation}
\overline{F_{12}}=
\left|{a\over {\bf x}_{12}}\right|^\kappa=
\left|{a\over {\bf x}_{12}}\right|^{2\pi\bar K+\delta\kappa}
\end{equation} 
in powers of $\delta\kappa$, then
the scaling exponent $\kappa$ governing the 
algebraic decay of $\overline{F_{12}}$
is given by
\begin{equation}
\kappa = 
2\pi \bar K - 
8\pi^4\left(\bar K^2- K^4g^2_A\right)
\left(\int_1^{{L\over a}}\! d x\ x^{3-2\pi\bar K}\right)\ 
\left(a^4Y_1^2\right)+{\cal O}(a^8Y_1^4).
\end{equation}
We see that the condition that the scaling exponent for the algebraic decay
of the two-point function be analytic in the fugacity $Y_1$ in the vicinity
of vanishing fugacity yields the parabolic boundary in Fig. 1. 

To fourth order in $Y_1\propto h_1/2t$ we must distinguish between
three different effects present {\it both with or without disorder}.
Indeed, we can write
\begin{equation}
F^{(4)}_{12}= A_{12}+B_{12}+C_{12}.
\end{equation}
Renormalizations of the interaction between two
external vortices due to:
i) four-body effects,
ii) three-body effects,
iii) two-body effects,
are denoted by $A_{12}$, $B_{12}$, and $C_{12}$, respectively.
Three-body and two-body effects must be accounted for when the separation
$|{\bf x}_{12}|$ between the two external charges is much larger than 
the hardcore radius $a$, and when the separation between vortices
is within the hardcore radius. Formally, we implement these
renormalization effects whenever coordinates of vortices coincide
in the integrands. This is never allowed due to the hardcore constraint 
for both $\overline{f^{(4)}_{12}}$ and $\overline{f^{(0)}_{12}Z^{(4)}}$.
However, coordinates can coincide when expanding the inverse
partition function in 
Eq. (\ref{power expansion in fugacity of two-point function}).
For example,
\begin{eqnarray}
\overline{f^{(2)}_{12}Z^{(2)}}&&=
\int \underbrace{d^2{\bf y}_1d^2{\bf y}_3}_{\neq}
\int \underbrace{d^2{\bf y}_2d^2{\bf y}_4}_{\neq}
\overline{
\left\langle
e^{{\rm i}
[
\chi({\bf y}_1)-
\chi({\bf y}_3)+ 
\chi({\bf x}_1)-
\chi({\bf x}_2)
]}
\right\rangle^{\rm unnor}_{h_1=0}\times
\left\langle
e^{{\rm i}
[
\chi({\bf y}_2)-
\chi({\bf y}_4) 
]}
\right\rangle^{\rm unnor}_{h_1=0}
}
\nonumber\\
&&=
\underbrace{\int_{1234}}_{\neq}
\overline{
\left\langle
e^{{\rm i}
[
\chi({\bf y}_1)-
\chi({\bf y}_3)+
\chi({\bf x}_1)-
\chi({\bf x}_2) 
]}
\right\rangle^{\rm unnor}_{h_1=0}
\times
\left\langle
e^{{\rm i}
[
\chi({\bf y}_2)-
\chi({\bf y}_4)
]}
\right\rangle^{\rm unnor}_{h_1=0}
}
\nonumber\\
&&+
\underbrace{\int_{134}}_{\neq}
\overline{
\left\langle
e^{{\rm i}
[
\chi({\bf y}_1)-
\chi({\bf y}_3)+
\chi({\bf x}_1)-
\chi({\bf x}_2) 
]}
\right\rangle^{\rm unnor}_{h_1=0}
\times
\left\langle
e^{{\rm i}
[
\chi({\bf y}_1)-
\chi({\bf y}_4)
]}
\right\rangle^{\rm unnor}_{h_1=0}
}
\nonumber\\
&&+
\underbrace{\int_{132}}_{\neq}
\overline{
\left\langle
e^{{\rm i}
[
\chi({\bf y}_1)-
\chi({\bf y}_3)+
\chi({\bf x}_1)-
\chi({\bf x}_2) 
]}
\right\rangle^{\rm unnor}_{h_1=0}
\times
\left\langle
e^{{\rm i}
[
\chi({\bf y}_2)-
\chi({\bf y}_1)
]}
\right\rangle^{\rm unnor}_{h_1=0}
}
\nonumber\\
&&+
\underbrace{\int_{134}}_{\neq}
\overline{
\left\langle
e^{{\rm i}
[
\chi({\bf y}_1)-
\chi({\bf y}_3)+
\chi({\bf x}_1)-
\chi({\bf x}_2) 
]}
\right\rangle^{\rm unnor}_{h_1=0}
\times
\left\langle
e^{{\rm i}
[
\chi({\bf y}_3)-
\chi({\bf y}_4)
]}
\right\rangle^{\rm unnor}_{h_1=0}
}
\nonumber\\
&&+
\underbrace{\int_{132}}_{\neq}
\overline{
\left\langle
e^{{\rm i}
[
\chi({\bf y}_1)-
\chi({\bf y}_3)+
\chi({\bf x}_1)-
\chi({\bf x}_2) 
]}
\right\rangle^{\rm unnor}_{h_1=0}
\times
\left\langle
e^{{\rm i}
[
\chi({\bf y}_2)-
\chi({\bf y}_3)
]}
\right\rangle^{\rm unnor}_{h_1=0}
}
\nonumber\\
&&+
\underbrace{\int_{13}}_{\neq}
\overline{
\left\langle
e^{{\rm i}
[
\chi({\bf y}_1)-
\chi({\bf y}_3)+
\chi({\bf x}_1)-
\chi({\bf x}_2) 
]}
\right\rangle^{\rm unnor}_{h_1=0}
\times
\left\langle
e^{{\rm i}
[
\chi({\bf y}_1)-
\chi({\bf y}_3)
]}
\right\rangle^{\rm unnor}_{h_1=0}
}
\nonumber\\
&&+
\underbrace{\int_{13}}_{\neq}
\overline{
\left\langle
e^{{\rm i}
[
\chi({\bf y}_1)-
\chi({\bf y}_3)+
\chi({\bf x}_1)-
\chi({\bf x}_2) 
]}
\right\rangle^{\rm unnor}_{h_1=0}
\times
\left\langle
e^{{\rm i}
[
\chi({\bf y}_3)-
\chi({\bf y}_1)
]}
\right\rangle^{\rm unnor}_{h_1=0}
}.
\end{eqnarray}

Four-body renormalization effects are
\begin{eqnarray}
&&
\overline{A_{12}}=
\\
&&
{1\over(2!)^2}
\underbrace{\int_{1234}}_{\neq}
\left[
{a\over|{\bf x}_{12}|}\times
{
a^2           |{\bf y}_{12}||{\bf y}_{34}|             \over
|{\bf y}_{13}||{\bf y}_{14}||{\bf y}_{23}||{\bf y}_{24}|
}\times
{
|{\bf y}_1-{\bf x}_1|
|{\bf y}_2-{\bf x}_1|
|{\bf y}_3-{\bf x}_2|
|{\bf y}_4-{\bf x}_2|
\over
|{\bf y}_1-{\bf x}_2|
|{\bf y}_2-{\bf x}_2|
|{\bf y}_3-{\bf x}_1|
|{\bf y}_4-{\bf x}_1|
}
\right]^{2\pi\bar K}-
\nonumber\\
&&
{1\over(2!)^2}\underbrace{\int_{3412}}_{\neq}
\left[
{a\over|{\bf x}_{12}|}
{
a^2           |{\bf y}_{12}||{\bf y}_{34}|             \over
|{\bf y}_{13}||{\bf y}_{14}||{\bf y}_{23}||{\bf y}_{24}|
}
\right]^{2\pi\bar K}
\left[
{
|{\bf y}_1-{\bf x}_1|
|{\bf y}_2-{\bf x}_1|
|{\bf y}_3-{\bf x}_2|
|{\bf y}_4-{\bf x}_2|
\over
|{\bf y}_1-{\bf x}_2|
|{\bf y}_2-{\bf x}_2|
|{\bf y}_3-{\bf x}_1|
|{\bf y}_4-{\bf x}_1|
}
\right]^{+2\pi K^2 g^{\ }_A}+
\nonumber\\
&&
\underbrace{\int_{3412}}_{\neq}
\left[
{a\over|{\bf x}_{12}|}
{
a^2
\over
|{\bf y}_{13}||{\bf y}_{24}|
}
\right]^{2\pi\bar K}
\!\!\!\times
\left[
{
|{\bf y}_{12}||{\bf y}_{34}|
\over
|{\bf y}_{14}||{\bf y}_{32}|
}
\times
{
|{\bf y}_1-{\bf x}_1|
|{\bf y}_2-{\bf x}_1|
|{\bf y}_3-{\bf x}_2|
|{\bf y}_4-{\bf x}_2|
\over
|{\bf y}_1-{\bf x}_2|
|{\bf y}_2-{\bf x}_2|
|{\bf y}_3-{\bf x}_1|
|{\bf y}_4-{\bf x}_1|
}
\right]^{+2\pi K^2 g^{\ }_A}-
\nonumber\\
&&
\underbrace{\int_{1432}}_{\neq}
\left[
{a\over|{\bf x}_{12}|}
{a^2\over|{\bf y}_{13}||{\bf y}_{24}|}
{
|{\bf y}_1-{\bf x}_1|
|{\bf y}_3-{\bf x}_2|
\over
|{\bf y}_1-{\bf x}_2|
|{\bf y}_3-{\bf x}_1|
}
\right]^{2\pi\bar K}
\left[
{|{\bf y}_{12}||{\bf y}_{34}|\over
|{\bf y}_{14}||{\bf y}_{32}|}
{
|{\bf y}_2-{\bf x}_1|
|{\bf y}_4-{\bf x}_2|
\over
|{\bf y}_2-{\bf x}_2|
|{\bf y}_4-{\bf x}_1|
}
\right]^{+2\pi K^2 g^{\ }_A}.
\nonumber
\end{eqnarray}
Assuming screening by a dilute gas of dipoles reduces the integration over the
coordinates ${\bf y}_1$, $\cdots$, ${\bf y}_4$ of the four thermal vortices to 
\begin{eqnarray}
\overline{A_{12}}&&\approx
2\times{1\over(2!)^2}
\left|{a^2\over{\bf x}_{12}}\right|^{2\pi\bar K}
\left\{
\underbrace{\int_{12}}_{\neq}
{
{\cal K}_{{\bf x}_1{\bf x}_2}({\bf y}_1,{\bf y}_2;2\pi\bar K)-
{\cal K}_{{\bf x}_1{\bf x}_2}({\bf y}_1,{\bf y}_2;2\pi K^2 g^{\ }_A)
\over
|{\bf y}_1-{\bf y}_2|^{2\pi\bar K}
}
\right\}^2
\nonumber\\
&&=
{1\over2}
\left|{a\over{\bf x}_{12}}\right|^{2\pi\bar K}
\left\{
8\pi^4\!\left(\bar K^2- K^4g^2_A\right)\!\!
\left(\int_a^{L}\!\!\! d|{\bf y}_{12}|\ 
|{\bf y}_{12}|^3\left|{{\bf y}_{12}\over a}\right|^{-2\pi\bar K}\right)
\ln\left|{{\bf x}_{12}\over a}\right|
\right\}^2.
\label{A_{12}bis}
\end{eqnarray}

Three-body renormalization effects are
\begin{eqnarray}
&&
\overline{B_{12}}=
\\
&&
2\underbrace{\int_{134}}_{\neq}
\left[
{a\over|{\bf x}_{12}|}
\right]^{2\pi\bar K}
\!\!\!\times
\left[
{
a^2
\over
|{\bf y}_{13}||{\bf y}_{14}|
}
\right]^{2\pi K} 
\!\!\!\times
\left[
{
a^4|{\bf y}_{34}|
\over
|{\bf y}_{14}|^2|{\bf y}_{13}|^2
}
\times
{
|{\bf y}_1-{\bf x}_1|^2
|{\bf y}_3-{\bf x}_2|
|{\bf y}_4-{\bf x}_2|
\over
|{\bf y}_1-{\bf x}_2|^2
|{\bf y}_3-{\bf x}_1|
|{\bf y}_4-{\bf x}_1|
}
\right]^{-2\pi K^2 g^{\ }_A}-
\nonumber\\
&&
2\underbrace{\int_{134}}_{\neq}
\left[
{a\over|{\bf x}_{12}|}
\right]^{2\pi\bar K}
\!\!\!\times
\left[
{
a^2
|{\bf y}_1-{\bf x}_1||{\bf y}_3-{\bf x}_2|
\over
|{\bf y}_{13}|
|{\bf y}_1-{\bf x}_2||{\bf y}_3-{\bf x}_1|
|{\bf y}_{14}|
}
\right]^{2\pi K}
\times
\nonumber\\
&&\hskip 3 true cm
\left[
{
a^4
|{\bf y}_{34}|
\over
|{\bf y}_{14}|^2|{\bf y}_{13}|^2
}
\times
{
|{\bf y}_1-{\bf x}_1|^2
|{\bf y}_3-{\bf x}_2|
|{\bf y}_4-{\bf x}_2|
\over
|{\bf y}_1-{\bf x}_2|^2
|{\bf y}_3-{\bf x}_1|
|{\bf y}_4-{\bf x}_1|
}
\right]^{-2\pi K^2 g^{\ }_A}+
\nonumber\\
&&
2\underbrace{\int_{132}}_{\neq}
\left[
{a\over|{\bf x}_{12}|}
\right]^{2\pi\bar K}
\!\!\!\times
\left[
{
a^2
\over
|{\bf y}_{13}|
|{\bf y}_{21}|
}
\right]^{2\pi K}
\!\!\!\times
\left[
{
a
\over
|{\bf y}_{32}|
}
\times
{
|{\bf y}_2-{\bf x}_1|
|{\bf y}_3-{\bf x}_2|
\over
|{\bf y}_2-{\bf x}_2|
|{\bf y}_3-{\bf x}_1|
}
\right]^{-2\pi K^2 g^{\ }_A}-
\nonumber\\
&&
2\underbrace{\int_{132}}_{\neq}
\left[
{a\over|{\bf x}_{12}|}
\right]^{2\pi\bar K}
\!\!\!\times
\left[
{
a^2
|{\bf y}_1-{\bf x}_1||{\bf y}_3-{\bf x}_2|
\over
|{\bf y}_{13}|
|{\bf y}_1-{\bf x}_2||{\bf y}_3-{\bf x}_1|
|{\bf y}_{21}|
}
\right]^{2\pi K}
\!\!\!\times
\left[
{
a
\over
|{\bf y}_{32}|
}
\times
{
|{\bf y}_2-{\bf x}_1|
|{\bf y}_3-{\bf x}_2|
\over
|{\bf y}_2-{\bf x}_2|
|{\bf y}_3-{\bf x}_1|
}
\right]^{-2\pi K^2 g^{\ }_A}.
\nonumber
\end{eqnarray}
After introducing the center of mass coordinates,
\begin{eqnarray}
&&
{\bf Y}:= {1\over3}\left({\bf y}_1+{\bf y}_2+{\bf y}_3\right),
\\
&&
{\bf y}_{12}:= {\bf y}_1-{\bf y}_2,
\\
&&
{\bf y}_{13}:= {\bf y}_1-{\bf y}_3,
\end{eqnarray}
it is possible to show that
\begin{eqnarray}
\overline{B_{12}}\left({h_1\over2t}\right)^4&&\approx
\left|{a\over{\bf x}_{12}}\right|^{2\pi\bar K}
\ln\left|{{\bf x}_{12}\over a}\right|
(2\pi)^3\times2\left(K^4g^2_A-\bar K^2\right)
Y_{(+1,+1;+2)}Y_{(-1,0;-1)}Y_{(0,-1;-1)}
{\tilde S^{(11)}\over2}
\nonumber\\
&&+
\left|{a\over{\bf x}_{12}}\right|^{2\pi\bar K}
\ln\left|{{\bf x}_{12}\over a}\right|
(2\pi)^3\times2\left(K^4g^2_A-\bar K^2\right)
Y_{(+1,+1;+2)}Y_{(-1,0;-1)}Y_{(0,-1;-1)}
{\tilde S^{(21)}\over2}
\nonumber\\
&&+
\left|{a\over{\bf x}_{12}}\right|^{2\pi\bar K}
\ln\left|{{\bf x}_{12}\over a}\right|
(2\pi)^3\times4K^3g^{\ }_A
Y_{(+1,-1;0)}Y_{(-1,0;-1)}Y_{(0,+1;+1)}
{\tilde S^{(13)}\over2}
\nonumber\\
&&-
\left|{a\over{\bf x}_{12}}\right|^{2\pi\bar K}
\ln\left|{{\bf x}_{12}\over a}\right|
(2\pi)^3\times4K^3g^{\ }_A
Y_{(+1,-1;0)}Y_{(-1,0;-1)}Y_{(0,+1;+1)}
{\tilde S^{(23)}\over2},
\label{ mean B_ 12 }
\end{eqnarray}
where we have defined four dimensionless integrals
\begin{eqnarray}
&&
\tilde S^{(11)}:=\!\!
\int\limits_{|{\bf y}_{12}|>a}\!\! {d^2{\bf y}_{12}\over a^2}\!\!
\int\limits_{|{\bf y}_{13}|>a}\!\! {d^2{\bf y}_{13}\over a^2}
\left|{a\over{\bf y}_{12}}\right|^{2\pi\overline{K(2)/2}}
\left|{{\bf y}_{13}-{\bf y}_{12}\over a}\right|^{-2\pi K^2g^{\ }_A}
\left|{a\over{\bf y}_{13}}\right|^{2\pi\overline{K(2)/2}}
{{\bf y}_{12}\cdot{\bf y}_{12}\over a^2},
\label{tilde S11}
\\
&&
\tilde S^{(13)}:=\!\!
\int\limits_{|{\bf y}_{12}|>a}\!\!  {d^2{\bf y}_{12}\over a^2}\!\!
\int\limits_{|{\bf y}_{13}|>a}\!\!  {d^2{\bf y}_{13}\over a^2}
\left|{a\over{\bf y}_{12}}\right|^{2\pi\overline{K(2)/2}}
\left|{{\bf y}_{13}-{\bf y}_{12}\over a}\right|^{-2\pi K^2g^{\ }_A}
\left|{a\over{\bf y}_{13}}\right|^{2\pi\overline{K(2)/2}}
{{\bf y}_{12}\cdot{\bf y}_{13}\over a^2},
\label{tilde S13}
\\
&&
\tilde S^{(21)}:=
\int\limits_{|{\bf y}_{12}|>a} {d^2{\bf y}_{12}\over a^2}
\int\limits_{|{\bf y}_{13}|>a} {d^2{\bf y}_{13}\over a^2}
\left|{a\over{\bf y}_{12}}\right|^{2\pi K}
\left|{{\bf y}_{13}-{\bf y}_{12}\over a}\right|^{+2\pi K^2g^{\ }_A}
\left|{a\over{\bf y}_{13}}\right|^{2\pi K}\
{{\bf y}_{12}\cdot{\bf y}_{12}\over a^2},
\label{tilde S21}
\\
&&
\tilde S^{(23)}:=
\int\limits_{|{\bf y}_{12}|>a}\!\! {d^2{\bf y}_{12}\over a^2}
\int\limits_{|{\bf y}_{13}|>a}\!\! {d^2{\bf y}_{13}\over a^2}
\left|{a\over{\bf y}_{12}}\right|^{2\pi K}
\left|{{\bf y}_{13}-{\bf y}_{12}\over a}\right|^{+2\pi K^2g^{\ }_A}
\left|{a\over{\bf y}_{13}}\right|^{2\pi K}\
{{\bf y}_{12}\cdot{\bf y}_{13}\over a^2}.
\label{tilde S23}
\end{eqnarray}
Here, we have introduced the dimensionless fugacities
$Y_{(\varepsilon_1,\varepsilon_2;\varepsilon_1+\varepsilon_2)}$,
$Y_{(0,\varepsilon;\varepsilon)}$, and $Y_{(\varepsilon,0;\varepsilon)}$,
where $\varepsilon_1=\varepsilon_2=\pm1$ and $\varepsilon=\pm1$,
respectively.
Their bare values are given by
$\left({a^2h_1\over2t}\right)^{4/3}$.
Whereas under a RG rescaling $a=e^la'$ with $0<l\ll1$, the $Y$'s
and product thereof are renormalized multiplicatively,
an open problem is to find numerical factors entering
additive renormalization effect due to three-body effects. 
This is difficult in this real space RG approach
since we must perform four-dimensional integrals with complicated 
integrands and intricate boundaries.

Form invariance under an infinitesimal
rescaling of the short distance cutoff $a$:
$a':= ae^l$, $0<l\ll1$,
of $\overline{B_{12}}$ implies that the three fugacities
($\varepsilon_1,\varepsilon_2=\pm1$)
\begin{equation} 
Y_{( \varepsilon_1, \varepsilon_2; \varepsilon_1+\varepsilon_2)},\quad
Y_{(-\varepsilon_1,             0;-\varepsilon_1              )},\quad
Y_{(             0,-\varepsilon_2;              -\varepsilon_2)},
\end{equation}
obey:
\begin{enumerate}
\item
They always appear in the combinations
\begin{equation}
Y_{( \varepsilon_1, \varepsilon_2; \varepsilon_1+\varepsilon_2)}\times
Y_{(-\varepsilon_1,             0;-\varepsilon_1              )}\times
Y_{(             0,-\varepsilon_2;              -\varepsilon_2)}.
\end{equation}
\item
Those combinations renormalize multiplicatively according to two rules
($\varepsilon=\pm1$)
\begin{eqnarray}
&&
\left[
Y_{( \varepsilon, \varepsilon; 2\varepsilon)}
Y_{(-\varepsilon,           0;- \varepsilon)}
Y_{(           0,-\varepsilon;- \varepsilon)}
\right]'=
\label{multiplicative rule 1}
Y_{( \varepsilon, \varepsilon; 2\varepsilon)}
Y_{(-\varepsilon,           0;- \varepsilon)}
Y_{(           0,-\varepsilon;- \varepsilon)}\
e^{6-2\pi\overline{K(2)}-2\pi K^2 g^{\ }_A},
\nonumber\\
&&
\left[
Y_{( \varepsilon,-\varepsilon;            0)}
Y_{(-\varepsilon,           0;- \varepsilon)}
Y_{(           0,+\varepsilon;+ \varepsilon)}
\right]'=
\label{multiplicative rule 2}
Y_{( \varepsilon,-\varepsilon;            0)}
Y_{(-\varepsilon,           0;- \varepsilon)}
Y_{(           0,+\varepsilon;+ \varepsilon)}\
e^{6-4\pi K+2\pi K^2 g^{\ }_A}.
\nonumber
\end{eqnarray}
\end{enumerate}
Additionally, if we require the individual multiplicative
renormalization rules
\begin{eqnarray}
&&
Y'    _{(\varepsilon,\varepsilon;2\varepsilon)}=
Y^{\ }_{(\varepsilon,\varepsilon;2\varepsilon)}
e^{l[2-\pi\overline{K(2)}]},
\\
&&
Y'    _{(\varepsilon,-\varepsilon;0)}=
Y^{\ }_{(\varepsilon,-\varepsilon;0)}
e^{l[2-2\pi K]},
\\
&&
Y'    _{(\varepsilon,0;\varepsilon)}=
Y^{\ }_{(\varepsilon,0;\varepsilon)}
e^{l[2-\pi\bar K)]},
\\
&&
Y'    _{(0,\varepsilon;\varepsilon)}=
Y^{\ }_{(0,\varepsilon;\varepsilon)}
e^{l[2-\pi\bar K]},
\end{eqnarray}
we can summarize 
Eqs. (\ref{multiplicative rule 1},\ref{multiplicative rule 2}) by
\begin{eqnarray}
&&
Y_{(+1,+1;2)}\times
Y_{(-1,0;-1)}\times
Y_{(0,-1;-1)}\sim
\label{three body fugacities  I}
e^{{\rm i}(\chi_1+\chi_2)({\bf y}_1)}\times
e^{-{\rm i}\chi_1({\bf y}_2)}\times
e^{-{\rm i}\chi_2({\bf y}_3)},
\\
&&
Y_{(+1,-1;0)}\times
Y_{(-1,0;-1)}\times
Y_{(0,+1;+1)}\sim
\label{three body fugacities II}
e^{{\rm i}(\chi_1-\chi_2)({\bf y}_1)}\times
e^{-{\rm i}\chi_1({\bf y}_2)}\times
e^{+{\rm i}\chi_2({\bf y}_3)}.
\end{eqnarray}
Equations (\ref{three body fugacities I},\ref{three body fugacities II})
tell us that any of the three thermal vortices 
renormalizing in $\overline{B_{12}}$
the CB interaction between two external vortices can be thought of as
some {\it local} linear combination 
of the two replicated real scalar fields $\chi_{1,2}$.
We remember that $\chi_{1,2}$
couple to the disorder in such a way that the fields $\chi'_{1,2}$
defined by
$\chi^{\ }_{1,2}=\chi'_{1,2}+{\rm i}2\pi K\theta$
are free scalar fields independent of the disorder $\theta$.
The fugacity $Y_{(\varepsilon_1,\varepsilon_2;\varepsilon_1+\varepsilon_2)}$
is thus labeled by three charges (measured in the appropriate units):
\begin{itemize}

\item
The thermal charge $\varepsilon_1$ of $\chi'_1$.

\item
The thermal charge $\varepsilon_2$ of $\chi'_2$.

\item
The disorder charge $\varepsilon_1+\varepsilon_2$ of $\theta$.

\end{itemize}

Finally, two-body renormalization effects are
\begin{eqnarray}
&&
\overline{C_{12}}=
\underbrace{\int_{13}}_{\neq}
\left[
{a\over|{\bf x}_{12}|}
\right]^{2\pi\bar K}\times
\\
&&
\Bigg\{
\left[
{
a
\over
|{\bf y}_{13}|
}
\right]^{2\pi\overline{K(2)}}
\!\!\!\times
\left[
{
|{\bf y}_1-{\bf x}_1|
|{\bf y}_3-{\bf x}_2|
\over
|{\bf y}_1-{\bf x}_2|
|{\bf y}_3-{\bf x}_1|
}
\right]^{-2\times2\pi K^2 g^{\ }_A}
+
\left[
{
a
\over
|{\bf y}_{13}|
}
\right]^{2\times2\pi K}-
\nonumber\\
&&
\left[
{
a
\over
|{\bf y}_{13}|
}
\right]^{2\pi\overline{K(2)}}
\!\!\!\times
\left[
{
|{\bf y}_1-{\bf x}_1|
|{\bf y}_3-{\bf x}_2|
\over
|{\bf y}_1-{\bf x}_2|
|{\bf y}_3-{\bf x}_1|
}
\right]^{2\pi K-2\times2\pi K^2 g^{\ }_A}
-
\left[
{
a
\over
|{\bf y}_{13}|
}
\right]^{2\times2\pi K}
\!\!\!\times
\left[
{
|{\bf y}_1-{\bf x}_1|
|{\bf y}_3-{\bf x}_2|
\over
|{\bf y}_1-{\bf x}_2|
|{\bf y}_3-{\bf x}_1|
}
\right]^{2\pi K}
\Bigg\}.
\nonumber
\end{eqnarray}
The major difference between $\overline{C_{12}}$ 
and Eq. (\ref{F_{12}^{(2)}}) is the exponent
of the coordinate ${\bf y}_{13}$ given by
$\overline {K(2)}= 2 K- 4 K^2 g_A$.
It is possible to show that
\begin{eqnarray}
\overline{C_{12}}\left({h_1\over2t}\right)^4&&\approx
\left|
{a\over{\bf x}_{12}}
\right|^{2\pi\bar K}
\ln\left|{{\bf x}_{12}\over a}\right|
8\pi^4\left\{4K^4g^2_A-{\left[\overline{K(2)}\right]^2\over4}\right\}
Y^2_{(1,1;2)}
\int_1^{{L\over a}} dy y^{3-2\pi\overline{K(2)}}
\nonumber\\
&&-
\left|{a\over{\bf x}_{12}}\right|^{2\pi\bar K}
\ln\left|{{\bf x}_{12}\over a}\right|
8\pi^4 K^2
Y^2_{(1,-1;0)}
\int_1^{{L\over a}} dy y^{3-4\pi K}.
\end{eqnarray}
Here, we have introduced the dimensionless fugacities  
$Y_{(1,1;2)}$ and $Y_{(1,-1;0)}$
whose  bare values are equal to
$\left({a^2h_1\over2t}\right)^2$.

In the absence of disorder, renormalization effects due to higher charge
vortices do not modify the boundary extracted from Eq. (\ref{F_{12}^{(2)}}).
This is nothing but the statement that $\cos(\chi)$ 
is the most relevant operator of the family 
$
\cos(\chi_1)\times\cdots\times\cos(\chi_q),\cdots,
\cos(q\chi)
$, $q\in{\bf N}$,
along the Gaussian fixed line $1/K\geq0$, $g_A=0$, $Y_1=0$.
However, in the presence of the random
vector potential $\tilde\partial_\mu\theta$
the relevance of $\cos(\chi_1)\times\cdots\times\cos(\chi_q)$ 
increases with 
$q\in{\bf N}$.

\eleq


\end{multicols}

\end{document}